\documentclass[usenatbib]{mn2e}
\usepackage{amsmath,fleqn}

\usepackage{subfigure}
\usepackage[pdftex]{graphicx}

\newcommand\half{{\textstyle\frac{1}{2}}}

\newcommand\rmb{\mathrm{b}}

\newcommand\rmd{\mathrm{d}}
\newcommand\rme{\mathrm{e}}

\newcommand\rmi{\mathrm{i}}

\newcommand\rmD{\mathrm{D}}

\newcommand\f{\frac}
\newcommand\p{\partial}
\newcommand\cst{\mathrm{constant}}
\title[Local and global dynamics of eccentric discs]
{Local and global dynamics of eccentric astrophysical discs}

\author[Gordon I.\ Ogilvie and Adrian J.\ Barker]
{Gordon I.\ Ogilvie and Adrian J.\ Barker\\
Department of Applied Mathematics and Theoretical Physics,
University of Cambridge, Centre for Mathematical Sciences,\\
Wilberforce Road, Cambridge CB3 0WA}

\begin{document}

\maketitle

\label{firstpage}
 
\begin{abstract}
  We formulate a local dynamical model of an eccentric disc in which
  the dominant motion consists of elliptical Keplerian orbits.  The
  model is a generalization of the well known shearing sheet, and is
  suitable for both analytical and computational studies of the local
  dynamics of eccentric discs.  It is spatially homogeneous in the
  horizontal dimensions but has a time-dependent geometry that
  oscillates at the orbital frequency.  We show how certain averages
  of the stress tensor in the local model determine the large-scale
  evolution of the shape and mass distribution of the disc.  The
  simplest solutions of the local model are laminar flows consisting
  of a (generally nonlinear) vertical oscillation of the disc.
  Eccentric discs lack vertical hydrostatic equilibrium because of the
  variation of the vertical gravitational acceleration around the
  eccentric orbit, and in some cases because of the divergence of the
  orbital velocity field associated with an eccentricity gradient.  We
  discuss the properties of the laminar solutions, showing that they
  can exhibit extreme compressional behaviour for eccentricities
  greater than about $0.5$, especially in discs that behave
  isothermally.  We also derive the linear evolutionary equations for
  an eccentric disc that follow from the laminar flows in the absence
  of a shear viscosity.  In a companion paper we show that these
  solutions are linearly unstable and we determine the associated
  growth rates and unstable modes.
\end{abstract}


\section{Introduction}

\subsection{Astrophysical motivation}

Eccentric discs, in which the dominant motion consists of elliptical
Keplerian orbits, occur in a wide variety of astrophysical situations.
For example, an eccentric gaseous disc is formed directly when a star
(or a giant planet) evolves, through scattering or secular
interaction, on to an orbit that closely approaches the galactic
centre (or the host star), and is tidally disrupted
\citep[e.g.][]{1979Natur.280..214G,2011ApJ...732...74G}; such a
process might be responsible for the gas cloud G2 near Sgr A$^*$ in
the Galactic Centre \citep{2014ApJ...786L..12G}.

In an eccentric binary star, a circumstellar or circumbinary disc
acquires a forced eccentricity from the binary orbit via secular
gravitational interaction.  The importance of this for planet
formation in binary stars has been recognized
\citep{2008MNRAS.386..973P}.  Even if the binary orbit is circular,
certain mean-motion resonances can allow a free eccentricity of the
disc to grow, initially exponentially \citep{1991ApJ...381..259L}.
These effects can of course occur in non-stellar binaries such as
binary black holes with accretion discs \citep{2005ApJ...634..921A},
planet--satellite systems with planetary rings
\citep{1981ApJ...243.1062G,1983AJ.....88.1560B} and protoplanetary
systems \citep{2006A&A...447..369K}.  Whether planet--disc
interactions lead to eccentricity excitation
\citep{2003ApJ...585.1024G,2003ApJ...587..398O,2006ApJ...652.1698D}
depends on the dynamics of eccentric discs, because of the strong
coupling between the planet and the disc.

Even in the absence of an orbiting companion, a disc may become
eccentric through an instability of the circular state, such as
viscous overstability \citep{1978MNRAS.185..629K,2001MNRAS.325..231O}.
In contrast to the naive expectation that viscosity tends to
circularize a disc, viscous overstability may explain the eccentricity
of decretion discs formed around rapidly rotating Be stars
\citep[e.g.][]{2013AARv..21...69R}.

In eccentric binaries the forced eccentricity of the disc is locked to
that of the binary and may not be easily detectable.  However, discs
with a free eccentricity precess as a result of their pressure and any
gravitational influences that cause a departure from Keplerian motion.
This is the generally accepted explanation of the superhump phenomenon
in the SU UMa class of dwarf novae \citep[e.g.][]{1995Warner}, in
which the accretion disc expands sufficiently during superoutbursts to
encounter the 3:1 resonance with the binary orbit and becomes
eccentric.  (Some other systems exhibit steady accretion and permanent
superhumps.)  The elliptical outer rim of the disc in OY Car was
measured by \citet{1992AA...263..147H} through the variation of
eclipses of the hot spot.  [For a critical analysis of observational
evidence for eccentric discs in SU UMa stars from a particular
standpoint, see \citet{2009AcA....59...89S}.]  Recently,
\textit{Kepler} has been used to observe superoutbursts and superhumps
with much greater accuracy
\citep{2013PASJ...65...97K,2013PASJ...65...50O}.  In dwarf novae the
optical emission is modulated at the frequency at which the disc
precesses in the frame that rotates with the binary orbit, owing to
the interaction between the eccentric mode and the tidal deformation.
Related phenomena are also reported in low-mass X-ray binaries,
although the radiative mechanisms are different
\citep{2001MNRAS.321..475H}.

\subsection{Theoretical and computational background}

Several theoretical and computational approaches have been taken to
the study of eccentric discs.  One is to try to generalize the
classical theory of viscous accretion discs to allow for orbits of
arbitrary eccentricity
\citep{1992MNRAS.255...92S,1994MNRAS.266..583L,2001MNRAS.325..231O}.
These analyses, of which the last is by far the most general, aim to
derive evolutionary equations for the shape and mass distribution of
eccentric discs due to viscous and other internal stresses.  Earlier,
equations governing the evolution of narrow and slightly eccentric
planetary rings were formulated by \citet{1983AJ.....88.1560B}.  A
separate body of theoretical work relates to eccentric collisionless
stellar discs in galactic nuclei, notably M31
\citep{1995AJ....110..628T,2003ApJ...599..237P}.

Small eccentricities are governed by linear equations which can be
derived through a perturbation analysis of a circular disc.  This
approach has been taken by, e.g., \citet{1983PASJ...35..249K},
\citet{1999MNRAS.308..984L}, \citet{2001AJ....121.1776T},
\citet{2002A&A...388..615P}, \citet{2006MNRAS.368.1123G} and
\citet{2008MNRAS.388.1372O}.

The broad conclusion of this work is that eccentricity can propagate
through a disc by means of pressure and self-gravity, as a slow
one-armed density wave, while viscosity causes it to diffuse (except
in cases where it is excited by viscous overstability).  Differential
apsidal precession due to the rapid rotation of the central object,
relativistic effects, self-gravity of the disc or the presence of
orbiting companions can be important, as can three-dimensional effects
due to the vertical structure and oscillation of the disc.

Numerical simulations of eccentric discs have mainly been carried out
using smoothed particle hydrodynamics (SPH), which readily produces
eccentric discs in circular binary stars with mass ratios typical of
SU UMa stars
\citep{1988MNRAS.232...35W,1991ApJ...381..268L,1996MNRAS.279..402M,1998MNRAS.297..323M,2000MNRAS.314L...1M,2007MNRAS.378..785S}.
More recently, grid-based simulations have also found the development
of eccentric discs in the presence of a planetary
\citep{2006A&A...447..369K} or stellar
\citep{2008AA...487..671K,2009A&A...508.1493M,2012AA...539A..98M} companion.

\citet{2005A&A...432..743P} found that eccentric discs are
hydrodynamically unstable in the absence of viscosity.  The
instability is three-dimensional and takes the form of a parametric
resonance of inertial waves, as also occurs in tidally distorted discs
\citep{1993ApJ...406..596G} and in the classic elliptical instability
of flows with non-circular streamlines \citep{2002AnRFM..34...83K}.  A
related phenomenon occurs in warped discs \citep{2013MNRAS.433.2420O}.
\citet{2005A&A...432..757P} carried out numerical simulations of the
instability of eccentric discs in the absence of vertical gravity and
found that it led to subsonic turbulence.

\subsection{Plan of this paper}

This paper is organized as follows.  In Section~\ref{s:large} we
describe the geometry of an eccentric disc and recall the properties
of the orbital coordinates defined by \citet{2001MNRAS.325..231O}.  We
formulate the hydrodynamic equations in this coordinate system and
obtain the evolutionary equations for an eccentric disc in terms of
orbital averages of force and stress components.  In
Section~\ref{s:local} we derive a local model of an eccentric disc,
which will be useful for analytical and computational studies of
instabilities and turbulence in eccentric discs.  In
Section~\ref{s:laminar} we consider the simplest hydrodynamic
solutions of this local model, which are non-hydrostatic and
necessarily involve a vertical oscillation of the disc; we also
discuss the evolution of eccentric discs under this laminar dynamics.

In a companion paper \citep{2014BO} we use the local model to analyse
the linear hydrodynamic stability of an eccentric disc.

\section{Large-scale geometry and dynamics of an eccentric disc}
\label{s:large}

\subsection{Introduction}

In a thin astrophysical disc, the orbital motion is hypersonic and
fluid elements follow ballistic trajectories to a first approximation.
Around a spherical central mass, these trajectories are Keplerian
orbits, which can have eccentricity and inclination.  A general
Keplerian disc involves smoothly nested orbits of variable
eccentricity and inclination: it is both elliptical and warped.

The case of a warped disc composed of variably inclined but circular
orbits is more familiar and was treated recently by
\citet{2013MNRAS.433.2403O}.  In this paper we consider instead the
case of an eccentric disc composed of variably elliptical but coplanar
orbits.  The general case of a warped and eccentric disc remains for
future work.

\begin{figure*}
 \subfigure{\includegraphics[trim=0cm 0cm 0cm 0cm, clip=true,width=0.44\textwidth]{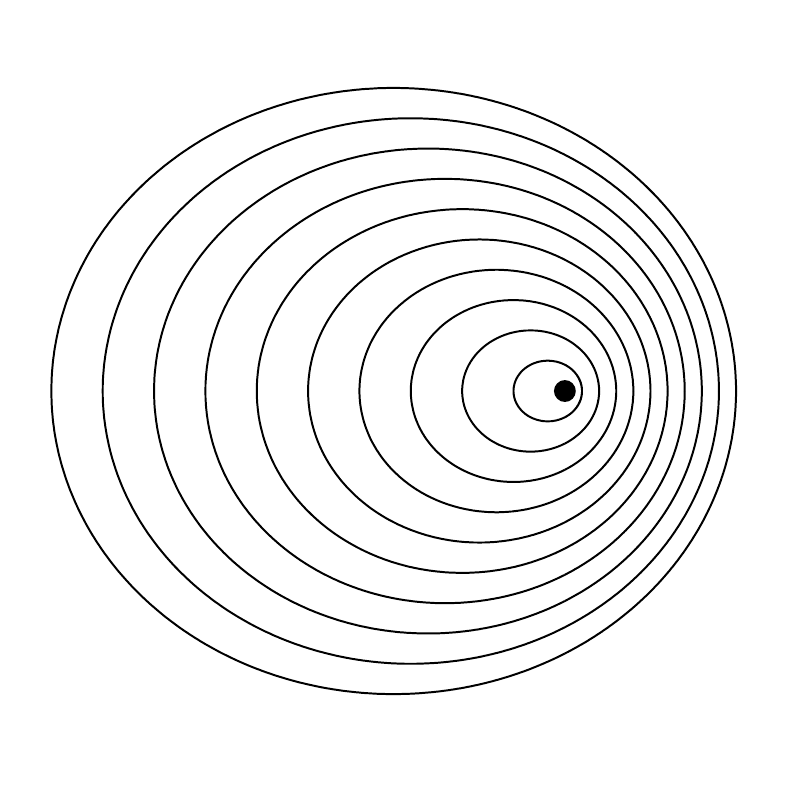} }
 \subfigure{\includegraphics[trim=0cm 0cm 0cm 0cm, clip=true,width=0.44\textwidth]{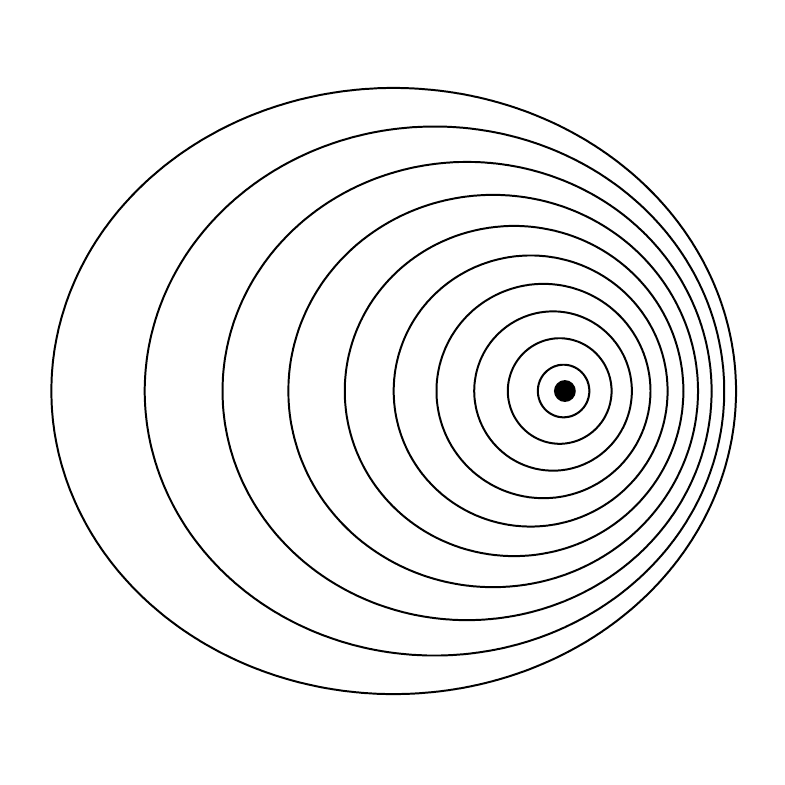} } \\
 \subfigure{\includegraphics[trim=0cm 0cm 0cm 0cm, clip=true,width=0.44\textwidth]{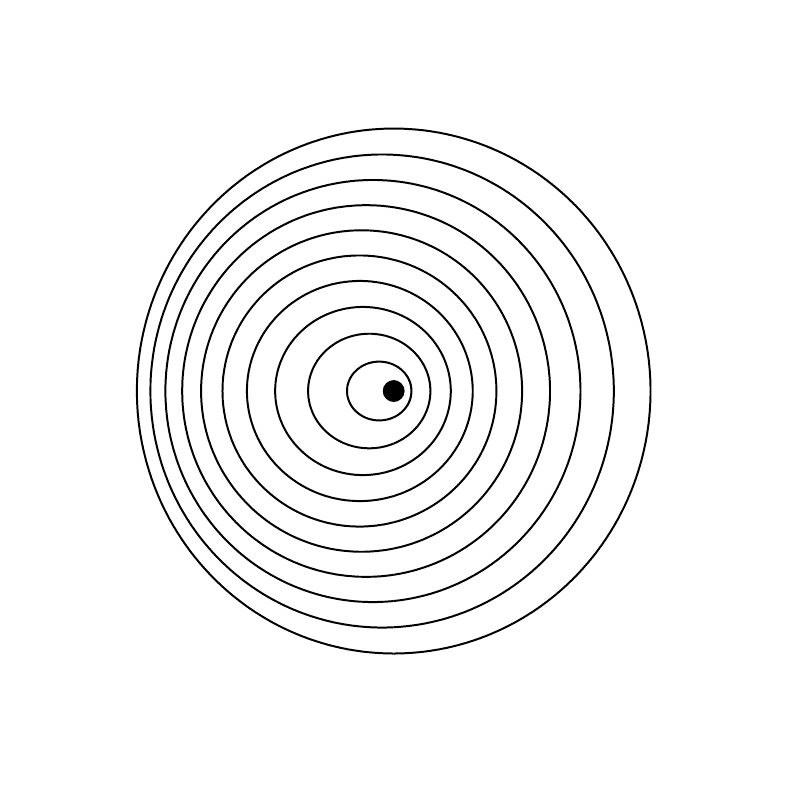} }
\subfigure{\includegraphics[trim=0cm 0cm 0cm 0cm, clip=true,width=0.44\textwidth]{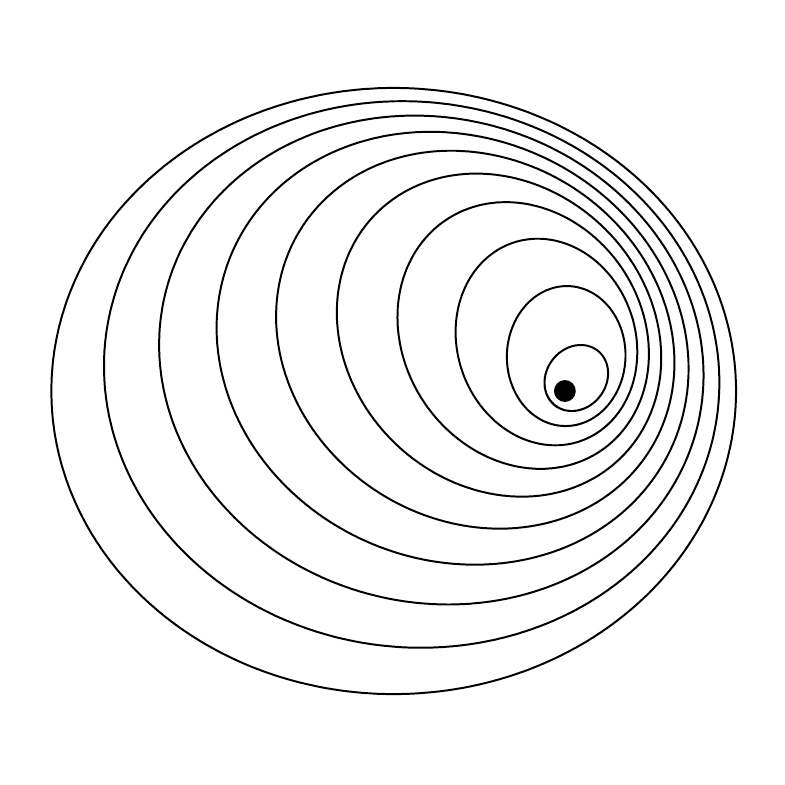} }
\caption{Four examples of eccentric discs.  Top left: uniformly
  eccentric disc with $e=0.5$ and $\omega=0$.  Top right: untwisted
  disc with $\omega=0$ and $e$ increasing linearly with $\lambda$ from
  $0$ to $0.5$.  Bottom left: untwisted disc with $\omega=0$ and $e$
  decreasing linearly with $\lambda$ from $0.5$ to $0$.  Bottom right:
  twisted disc with $e=0.5$ and $\omega$ increasing logarithmically
  with $\lambda$ such that $\lambda\omega'=1$.}
\label{f:disc}
\end{figure*}

The dominant motion in an eccentric disc is orbital motion in the form
of Keplerian ellipses.  Orbital precession due to a small departure of
the gravitational potential from that of a point mass, or due to weak
relativistic effects, can be treated, along with the collective
effects of the disc, as a perturbation of the Keplerian motion.  The
eccentric disc can therefore be considered, to a first approximation,
as a continuum of nested elliptical rings (Fig.~\ref{f:disc}), whose
shape can be regarded as fixed in a non-rotating frame on the
timescale of the orbital motion.

Following \citet{2001MNRAS.325..231O}, we label the orbits using their
semi-latus rectum $\lambda=a(1-e^2)$, where $a$ and $e$ are the
semi-major axis and the eccentricity.  The semi-latus rectum is
directly related to the specific angular momentum
$\ell=(GM\lambda)^{1/2}$, where $G$ is Newton's constant and $M$ is
the central mass.  In an eccentric disc, the eccentricity $e$ and the
longitude of pericentre $\omega$ can be regarded as functions of
$\lambda$, and can be conveniently combined into the \textit{complex
  eccentricity} $E=e\,\rme^{\rmi\omega}$.  (Note that the
eccentricity, $e$, and the base of natural logarithms, $\rme$, are
distinguished typographically.)  The real and imaginary parts of $E$
are equivalent to the components of the \textit{eccentricity vector}
$(e\cos\omega,e\sin\omega)$.

\subsection{Orbital coordinates}

\begin{figure}
\subfigure{\includegraphics[trim=0cm 0cm 0cm 0cm, clip=true,width=0.44\textwidth]{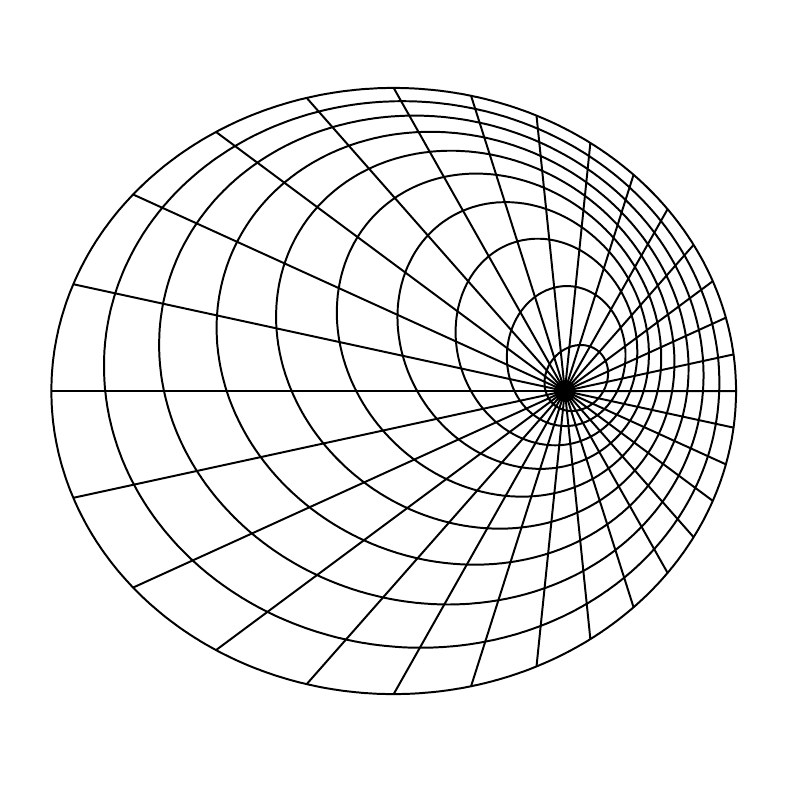} }
\caption{Orbital coordinates $(\lambda,\phi)$ in an example of an
  eccentric disc.  The semi-latus rectum $\lambda$ is used to label
  the orbits and has the role of a quasi-radial coordinate.  The
  second coordinate is the usual azimuthal angle $\phi$.}
\label{f:orbital}
\end{figure}

Again following \citet{2001MNRAS.325..231O}, we make use of
\textit{orbital coordinates} $(\lambda,\phi)$ in the plane of the disc
(Fig.~\ref{f:orbital}), instead of polar coordinates $(r,\phi)$.  The
transformation between them is given by the polar equation of an
ellipse,
\begin{equation}
  r=R(\lambda,\phi)=\f{\lambda}{1+e(\lambda)\cos[\phi-\omega(\lambda)]}.
\label{ellipse}
\end{equation}
The semi-latus rectum $\lambda$ can be thought of as a quasi-radial
coordinate that replaces $r$, while $\phi$ is the usual azimuthal
angular coordinate.

Orbital coordinates are, of course, well adapted to the geometry of an
eccentric disc, in which the Keplerian ellipses correspond to the
curves $\lambda=\cst$.  The disadvantage of these coordinates is that
they are not orthogonal.  Since we will need to carry out vector and
tensor calculus in these coordinates, we must first define a number of
geometrical quantities, quoting results from
\citet{2001MNRAS.325..231O}.

The components of the metric tensor $g_{ij}$ and of its inverse
$g^{ij}$ are
\begin{equation}
  g_{\lambda\lambda}=R_\lambda^2,\qquad
  g_{\lambda\phi}=g_{\phi\lambda}=R_\lambda R_\phi,\qquad
  g_{\phi\phi}=R^2+R_\phi^2,
\end{equation}
\begin{equation}
  g^{\lambda\lambda}=\f{R^2+R_\phi^2}{R^2R_\lambda^2},\qquad
  g^{\lambda\phi}=g^{\phi\lambda}=-\f{R_\phi}{R^2R_\lambda},\qquad
  g^{\phi\phi}=\f{1}{R^2},
\end{equation}
where the subscripts on $R$ denote partial derivatives of the function
$R(\lambda,\phi)$.  The components of the Levi-Civita connection
(Christoffel symbols) are
\begin{equation}
  \Gamma^\lambda_{\lambda\lambda}=\f{R_{\lambda\lambda}}{R_\lambda},\qquad
  \Gamma^\lambda_{\lambda\phi}=\Gamma^\lambda_{\phi\lambda}=\f{R_{\lambda\phi}}{R_\lambda}-\f{R_\phi}{R},
\end{equation}
\begin{equation}
  \Gamma^\lambda_{\phi\phi}=-\f{(R^2+2R_\phi^2-RR_{\phi\phi})}{RR_\lambda},
\end{equation}
\begin{equation}
  \Gamma^\phi_{\lambda\lambda}=0,\qquad
  \Gamma^\phi_{\lambda\phi}=\Gamma^\phi_{\phi\lambda}=\f{R_\lambda}{R},\qquad
  \Gamma^\phi_{\phi\phi}=\f{2R_\phi}{R}.
\end{equation}
The orbital coordinate system is easily extended to three dimensions
by adding the third coordinate $z$.  Apart from $g_{zz}=g^{zz}=1$, all
metric and connection components involving $z$ vanish.  The Jacobian
of the coordinate system is
\begin{equation}
  J=\f{\p(x,y,z)}{\p(\lambda,\phi,z)}=(\det g_{ij})^{1/2}=RR_\lambda.
\end{equation}
In order that $J>0$, i.e.\ that the orbits are closed and nested
without intersection, we require $|E|<1$ and $|E-\lambda E'|<1$, where
the prime denotes a derivative with respect to $\lambda$.  Therefore
both the eccentricity and the eccentricity gradient must be
sufficiently small.

Further useful geometrical relations are derived in
Appendix~\ref{s:relations}.

Although by convention $e\ge0$, the same orbit is obtained by
reversing the sign of $e$ and increasing (or decreasing) $\omega$ by
$\pi$, which leaves the complex eccentricity $E$ unchanged.  In
situations where $E(\lambda)$ has a simple zero corresponding to a
circular orbit, $E'$ is well defined and non-zero on that orbit.
However, as conventionally defined, $e$ has a discontinuous gradient
there and $\omega$ changes abruptly by $\pi$.  In order to avoid this
discontinuity and to construct a local model in such a case, we will
allow $e$ to become negative on one side of the circular orbit so that
$e'$ is continuous there.

In principle the orbital coordinates are time-dependent as well as
non-orthogonal.  The difficulties that this introduces were treated by
\citet{2001MNRAS.325..231O}.  In this paper we mainly circumvent these
difficulties by assuming that the orbital geometry is fixed on the
orbital timescale.  However, the slow time-dependence of the
coordinates is taken into account in deriving the evolutionary
equations for the eccentric disc in Section~\ref{s:evolution}.

\subsection{Hydrodynamic equations}

We consider an ideal fluid satisfying the equation of motion,
\begin{equation}
  (\p_t+u^j\nabla_j)u^i=-g^{ij}\left(\nabla_j\Phi+\f{1}{\rho}\nabla_jp\right),
\label{motion}
\end{equation}
the equation of mass conservation,
\begin{equation}
  \p_t\rho+\nabla_i(\rho u^i)=0,
\label{mass}
\end{equation}
and the thermal energy equation,
\begin{equation}
  \rmD s=0,
\label{entropy}
\end{equation}
where $u^i$ is the velocity, $\nabla_i$ is the covariant derivative,
$\Phi$ is the gravitational potential, $\rho$ is the density, $p$ is
the pressure, $s$ is the specific entropy and $\rmD=\p_t+u^i\p_i$ is
the Lagrangian derivative acting on scalar fields.  We assume that the
disc is of sufficiently low mass that self-gravity may be neglected.

We work with the contravariant velocity components $u^i$, which are
simply the rates of change of the orbital coordinates of fluid
elements.  Equation~(\ref{motion}) can be written using partial
derivatives as
\begin{equation}
  \rmD u^i+\Gamma^i_{jk}u^ju^k=-g^{ij}\left(\p_j\Phi+\f{1}{\rho}\p_jp\right),
\label{dui}
\end{equation}
while equation~(\ref{mass}) can be written either as
\begin{equation}
  \rmD\rho=-\f{\rho}{J}\p_i(Ju^i)
\label{drho}
\end{equation}
or in the conservative form
\begin{equation}
  \p_t(J\rho)+\p_i(J\rho u^i)=0.
\label{djrho}
\end{equation}
An alternative to equation~(\ref{entropy}) is
\begin{equation}
  \rmD p=-\f{\gamma p}{J}\p_i(Ju^i),
\label{dp}
\end{equation}
where $\gamma=(\p\ln p/\p\ln\rho)_s$ is the adiabatic index.

The conservative form of the equation for the total energy of the fluid is
\begin{equation}
  \p_t(J\rho\mathcal{E}_\mathrm{tot})+\p_i[J(\rho\mathcal{E}_\mathrm{tot}+p)u^i]=0,
\end{equation}
where
\begin{equation}
  \mathcal{E}_\mathrm{tot}=\half g_{ij}u^iu^j+\varepsilon+\Phi
\end{equation}
is the specific total energy and $\varepsilon$ is the specific
internal energy.  This result follows from equations (\ref{dui}) and
(\ref{djrho}) when the relations $\rmd\varepsilon=T\,\rmd
s-p\,\rmd(\rho^{-1})$ and
$\p_kg_{ij}=g_{il}\Gamma^l_{jk}+g_{lj}\Gamma^l_{ik}$ are used,
provided that the gravitational potential is independent of $t$.  In
detail, equations (\ref{dui}), (\ref{drho}) and (\ref{dp}) are
\begin{eqnarray}
  \lefteqn{\rmD u^\lambda+\Gamma^\lambda_{\lambda\lambda}(u^\lambda)^2+2\Gamma^\lambda_{\lambda\phi}u^\lambda u^\phi+\Gamma^\lambda_{\phi\phi}(u^\phi)^2}&\nonumber\\
  &&=-g^{\lambda\lambda}\left(\p_\lambda\Phi+\f{1}{\rho}\p_\lambda p\right)-g^{\lambda\phi}\left(\p_\phi\Phi+\f{1}{\rho}\p_\phi p\right),
\label{dul}
\end{eqnarray}
\begin{eqnarray}
  \lefteqn{\rmD u^\phi+2\Gamma^\phi_{\lambda\phi}u^\lambda u^\phi+\Gamma^\phi_{\phi\phi}(u^\phi)^2}&\nonumber\\
  &&=-g^{\lambda\phi}\left(\p_\lambda\Phi+\f{1}{\rho}\p_\lambda p\right)-g^{\phi\phi}\left(\p_\phi\Phi+\f{1}{\rho}\p_\phi p\right),
\label{dup}
\end{eqnarray}
\begin{equation}
  \rmD u^z=-\p_z\Phi-\f{1}{\rho}\p_zp,
\label{duz}
\end{equation}
\begin{equation}
  \rmD\rho=-\rho\left[\f{1}{J}\p_\lambda(Ju^\lambda)+\f{1}{J}\p_\phi(Ju^\phi)+\p_zu^z\right]
\label{durho}
\end{equation}
and
\begin{equation}
  \rmD p=-\gamma p\left[\f{1}{J}\p_\lambda(Ju^\lambda)+\f{1}{J}\p_\phi(Ju^\phi)+\p_zu^z\right],
\label{dupressure}
\end{equation}
with $\rmD=\p_t+u^\lambda\p_\lambda+u^\phi\p_\phi+u^z\p_z$.  The
alternative, conservative form of equation~(\ref{durho}) is
\begin{equation}
  \p_t(J\rho)+\p_\lambda(J\rho u^\lambda)+\p_\phi(J\rho u^\phi)+\p_z(J\rho u^z)=0.
\end{equation}
(In Section~\ref{s:evolution} we will use a modified form of this
equation that takes into account the slow time-dependence of the
orbital coordinates.)

\subsection{Orbital motion}
\label{s:orbital}

The orbital motion corresponds to the velocity field $u^i=\omega^i$,
where $\omega^\phi=\Omega(\lambda,\phi)$ is the orbital angular
velocity and $\omega^\lambda=\omega^z=0$.  This velocity field should
satisfy the equation of motion in the midplane $z=0$, where
$\Phi=\Phi_0(R)$, when the pressure is neglected, i.e.
\begin{equation}
  \Gamma^\lambda_{\phi\phi}\Omega^2=-g^{\lambda\lambda}\p_\lambda\Phi_0-g^{\lambda\phi}\p_\phi\Phi_0,
\label{orbital1}
\end{equation}
\begin{equation}
  \Omega\p_\phi\Omega+\Gamma^\phi_{\phi\phi}\Omega^2=-g^{\lambda\phi}\p_\lambda\Phi_0-g^{\phi\phi}\p_\phi\Phi_0.
\label{orbital2}
\end{equation}
Since $\Phi_0$ is a function of $R$ only, these equations simplify to
\begin{equation}
  R^2\Omega^2=\lambda\f{\rmd\Phi_0}{\rmd R},
\end{equation}
\begin{equation}
  \p_\phi(R^2\Omega)=0,
\end{equation}
and are satisfied, as expected, when $\Phi_0=-GM/R$ and
$R^2\Omega=\ell=(GM\lambda)^{1/2}$, i.e.\
\begin{equation}
  \Omega=\left(\f{GM}{\lambda^3}\right)^{1/2}[1+e\cos(\phi-\omega)]^2.
\end{equation}

\subsection{Evolution of mass, angular momentum and eccentricity}
\label{s:evolution}

A principal aim of a theory of eccentric discs is to obtain a system
of equations that govern the evolution of the shape and mass
distribution of the disc.  Unlike the case of a warped disc composed
of circular orbits, these equations do not follow simply from the
conservation of mass and angular momentum.  The evolution of the
complex eccentricity, or eccentricity vector, is more subtle and is
not purely conservative in nature.

We consider first the case of a test particle in a Keplerian orbit in
the plane $z=0$ and subject to a perturbing force within that plane.
Its motion is governed by
\begin{equation}
  \ddot r-\f{\ell^2}{r^3}=-\f{GM}{r^2}+f_r,
\label{fr}
\end{equation}
\begin{equation}
  \dot\ell=rf_\phi,
\label{fp}
\end{equation}
where $\ell=r^2\dot\phi$ is the specific angular momentum and $f_r$
and $f_\phi$ are the (orthogonal) polar components of the perturbing
force per unit mass.  The osculating orbital elements are defined by
equating the instantaneous position and velocity of the particle with
those of a Keplerian orbit.  Thus
\begin{equation}
  r=\f{\lambda}{1+e\cos\theta},\qquad
  \dot r=\f{\ell}{\lambda}e\sin\theta,
\end{equation}
where $\lambda=\ell^2/GM$ is the semi-latus rectum, $e$ is the
eccentricity, $\theta=\phi-\omega$ is the true anomaly and $\omega$ is
the longitude of pericentre.  From the above relations we have
\begin{equation}
  e\,\rme^{-\rmi\theta}=e\cos\theta-\rmi e\sin\theta=\f{\lambda}{r}-1-\f{\rmi\lambda\dot r}{\ell}
\end{equation}
and so
\begin{equation}
  E=e\,\rme^{\rmi\omega}=\left(\f{\lambda}{r}-1-\f{\rmi\lambda\dot r}{\ell}\right)\rme^{\rmi\phi},
\end{equation}
which therefore evolves according to
\begin{equation}
  \ell\dot E=rf_\phi(\rme^{\rmi\phi}+E)+\lambda(f_\phi-\rmi f_r)\,\rme^{\rmi\phi}.
\end{equation}
Although unfamiliar in this form, this equation is equivalent to the
Gauss perturbation equations of celestial mechanics in the case of
planar motion.

If instead we use the contravariant orbital components $f^\lambda$ and
$f^\phi$, which are related by $f_r=R_\lambda f^\lambda+R_\phi f^\phi$
and $f_\phi=Rf^\phi$, then we can write (using equation~\ref{rp})
\begin{equation}
  \dot\ell=R^2f^\phi,
\label{ldot}
\end{equation}
\begin{equation}
  \ell\dot E=2R^2f^\phi(\rme^{\rmi\phi}+E)-\rmi\lambda R_\lambda f^\lambda\,\rme^{\rmi\phi}.
\label{ledot}
\end{equation}

We now consider a continuous disc.  The equation of mass conservation
in a three-dimensional conservative form that takes into account the
time-dependence of the orbital coordinates is
\citep{2001MNRAS.325..231O}
\begin{equation}
  \p_t(J\rho)+\p_\lambda[J\rho(\dot\lambda+u^\lambda)]+\p_\phi(J\rho u^\phi)+\p_z(J\rho u^z)=0,
\end{equation}
where $\dot\lambda$ is the rate of change of $\lambda$ with time in an
inertial coordinate system, due to the slow evolution of the orbital
geometry.  Integrating this equation with respect to $\phi$ and $z$
over the full extent of the disc, and assuming that no mass is gained
or lost vertically, we obtain the one-dimensional conservative form
\begin{equation}
  \p_t\mathcal{M}+\p_\lambda\mathcal{F}=0,
\label{mass1}
\end{equation}
where
\begin{equation}
  \mathcal{M}=\iint J\rho\,\rmd\phi\,\rmd z=\int J\Sigma\,\rmd\phi
\end{equation}
is the one-dimensional mass density with respect to $\lambda$,
$\Sigma=\int\rho\,\rmd z$ being the surface density, and
\begin{equation}
  \mathcal{F}=\iint J\rho(\dot\lambda+u^\lambda)\,\rmd\phi\,\rmd z
\end{equation}
is the quasi-radial mass flux.  Note that the mass of the disc is
\begin{equation}
  \iiint J\rho\,\rmd\lambda\,\rmd\phi\,\rmd z=\int\mathcal{M}\,\rmd\lambda,
\end{equation}
where the integral is carried out over an appropriate range of
$\lambda$.  We can also write
\begin{equation}
  \p_t\mathcal{M}+\p_\lambda(\mathcal{M}\bar v^\lambda)=0,
\end{equation}
where $\bar v^\lambda=\mathcal{F}/\mathcal{M}$ is the mean
quasi-radial velocity.

Given that $\ell$ is independent of $\phi$ and $z$, the angular momentum
equation has the one-dimensional form
\begin{equation}
  \p_t(\mathcal{M}\ell)+\p_\lambda(\mathcal{F}\ell)=\iint J\rho R^2f^\phi\,\rmd\phi\,\rmd z,
\end{equation}
or, equivalently,
\begin{equation}
  \mathcal{M}(\p_t+\bar v^\lambda\p_\lambda)\ell=\iint J\rho R^2f^\phi\,\rmd\phi\,\rmd z,
\end{equation}
which is the continuum analogue of equation~(\ref{ldot}).

Let us consider the case of internal perturbing forces that are due to
stress divergences, i.e.\ $\rho f^i=\nabla_jT^{ij}$, where $T^{ij}$ is
a symmetric stress tensor describing the collective effects of the
disc (pressure, viscosity, self-gravity, etc.).  Then
\citep{2001MNRAS.325..231O}
\begin{eqnarray}
  \lefteqn{\rho f^\lambda=\f{1}{JR_\lambda}\p_\lambda(JR_\lambda T^{\lambda\lambda})+\f{R^2}{JR_\lambda^2}\p_\phi\left(\f{JR_\lambda^2}{R^2}T^{\lambda\phi}\right)}&\nonumber\\
  &&\qquad-\f{R^2}{\lambda R_\lambda}T^{\phi\phi}+\p_z T^{\lambda z},
\end{eqnarray}
\begin{equation}
  \rho f^\phi=\f{1}{JR^2}\p_\lambda(JR^2T^{\lambda\phi})+\f{1}{JR^2}\p_\phi(JR^2T^{\phi\phi})+\p_z T^{\phi z}.
\end{equation}
The terms in $J\rho R^2f^\phi$ involving $\p_\phi$ and $\p_z$
integrate to zero, assuming suitable boundary conditions in $z$ that
ensure that no angular momentum is lost or gained vertically, and it
follows that
\begin{equation}
  \p_t(\mathcal{M}\ell)+\p_\lambda(\mathcal{F}\ell)=\iint\p_\lambda(JR^2T^{\lambda\phi})\,\rmd\phi\,\rmd z,
\end{equation}
which can be written in the conservative form
\begin{equation}
  \p_t(\mathcal{M}\ell)+\p_\lambda(\mathcal{F}\ell+\mathcal{G})=0,
\end{equation}
where
\begin{equation}
  \mathcal{G}=-\iint JR^2T^{\lambda\phi}\,\rmd\phi\,\rmd z
\end{equation}
is the internal torque.  With the help of the equation of mass
conservation~(\ref{mass1}), and the fact that $\ell$ depends only on
$\lambda$, it simplifies to
\begin{equation}
  \mathcal{F}\,\f{\rmd\ell}{\rmd\lambda}+\f{\p\mathcal{G}}{\p\lambda}=0,
\end{equation}
which determines the mass flux $\mathcal{F}$ (or the mean quasi-radial
velocity $\bar v^\lambda=\mathcal{F}/\mathcal{M}$) instantaneously in
terms of the torque distribution.  As expected, the stress component
$T^{\lambda\phi}$ determines the redistribution of angular momentum
within the disc and thereby regulates the accretion flow.

The eccentricity equation is less obvious because it is not
conservative.  We can expect the continuum analogue of
equation~(\ref{ledot}) to be
\begin{eqnarray}
  \lefteqn{\ell\mathcal{M}(\p_t+\bar v^\lambda\p_\lambda)E}&\nonumber\\
  &&=\iint J\rho\left[2R^2f^\phi(\rme^{\rmi\phi}+E)-\rmi\lambda R_\lambda f^\lambda\,\rme^{\rmi\phi}\right]\rmd\phi\,\rmd z.
\end{eqnarray}
When $\rho f^i=\nabla_jT^{ij}$, the terms involving $\p_z$ again
integrate to zero.  The terms involving $\p_\phi$ are less obvious and
an integration by parts is needed to obtain
\begin{eqnarray}
  \lefteqn{\ell\mathcal{M}(\p_t+\bar v^\lambda\p_\lambda)E=\iint\bigg[2(\rme^{\rmi\phi}+E)\p_\lambda(JR^2T^{\lambda\phi})}&\nonumber\\
  &&\quad-\rmi\lambda\,\rme^{\rmi\phi}\p_\lambda(JR_\lambda T^{\lambda\lambda})-\rmi\,\rme^{\rmi\phi}JR^2T^{\phi\phi}\nonumber\\
  &&\quad+\rmi\lambda\f{JR_\lambda^2}{R^2}T^{\lambda\phi}\p_\phi\left(\f{R^2}{R_\lambda}\rme^{\rmi\phi}\right)\bigg]\,\rmd\phi\,\rmd z.
\end{eqnarray}
Using the relation~(\ref{rlp}) we obtain
\begin{eqnarray}
  \lefteqn{\ell\mathcal{M}(\p_t+\bar v^\lambda\p_\lambda)E=\iint\bigg[2(\rme^{\rmi\phi}+E)\p_\lambda(JR^2T^{\lambda\phi})}&\nonumber\\
  &&\quad-\rmi\lambda\,\rme^{\rmi\phi}\p_\lambda(JR_\lambda T^{\lambda\lambda})-\rmi\,\rme^{\rmi\phi}JR^2T^{\phi\phi}\nonumber\\
  &&\quad-\f{JR^2}{\lambda}(\rme^{\rmi\phi}+E-\lambda E')T^{\lambda\phi}\bigg]\,\rmd\phi\,\rmd z.
\end{eqnarray}
This result is consistent with equation~(167) of
\citet{2001MNRAS.325..231O}, which was derived more formally using the
method of multiple time-scales, and can be written in various ways
(especially concerning where to place the $\p_\lambda$).

In order to close the equations governing the shape and mass
distribution of the disc, the four stress integrals that are needed are
therefore
\begin{equation}
  \iint JR^2T^{\lambda\phi}\,\rmd\phi\,\rmd z,
\label{stress1}
\end{equation}
\begin{equation}
  \iint JR^2T^{\lambda\phi}\rme^{\rmi\phi}\,\rmd\phi\,\rmd z,
\end{equation}
\begin{equation}
  \iint JR_\lambda T^{\lambda\lambda}\rme^{\rmi\phi}\,\rmd\phi\,\rmd z,
\end{equation}
\begin{equation}
  \iint JR^2T^{\phi\phi}\rme^{\rmi\phi}\,\rmd\phi\,\rmd z.
\label{stress4}
\end{equation}

External forces acting on the disc, which would include the force due
to any departure of the gravitational potential from that of a point
mass, also contribute to the evolution of $\ell$ and $E$ in the
obvious way.  Thus the governing equations in the presence of internal
forces (described by $T^{ij}$) and external forces (described by
$f^i$) are
\begin{equation}
  \p_t\mathcal{M}+\p_\lambda(\mathcal{M}\bar v^\lambda)=0,
\end{equation}
\begin{equation}
  \mathcal{M}\bar v^\lambda\f{\rmd\ell}{\rmd\lambda}=\iint\p_\lambda(JR^2T^{\lambda\phi})\,\rmd\phi\,\rmd z+\iint J\rho R^2f^\phi\,\rmd\phi\,\rmd z,
\end{equation}
\begin{eqnarray}
  \lefteqn{\ell\mathcal{M}(\p_t+\bar v^\lambda\p_\lambda)E=\iint\bigg[2(\rme^{\rmi\phi}+E)\p_\lambda(JR^2T^{\lambda\phi})}&\nonumber\\
  &&\quad-\rmi\lambda\,\rme^{\rmi\phi}\p_\lambda(JR_\lambda T^{\lambda\lambda})-\rmi\,\rme^{\rmi\phi}JR^2T^{\phi\phi}\nonumber\\
  &&\quad-\f{JR^2}{\lambda}(\rme^{\rmi\phi}+E-\lambda E')T^{\lambda\phi}\bigg]\,\rmd\phi\,\rmd z\nonumber\\
  &&\quad+\iint J\rho\left[2R^2f^\phi(\rme^{\rmi\phi}+E)-\rmi\lambda R_\lambda f^\lambda\,\rme^{\rmi\phi}\right]\rmd\phi\,\rmd z.
\label{edot}
\end{eqnarray}

An important aspect of the three-dimensional theory of eccentric discs
is that the point-mass potential makes a non-zero contribution to
$f^\lambda$.  When the potential $\Phi=-GM(R^2+z^2)^{-1/2}$ is
expanded in a Taylor series about the midplane $z=0$ of a thin disc,
we obtain
\begin{equation}
  \Phi=\Phi_0(R)+\Phi_2(R)\half z^2+O(z^4),
\label{taylor}
\end{equation}
where $\Phi_0=-GM/R$ is the potential in the midplane, while
$\Phi_2=GM/R^3$ has a different dependence on $R$.  Its contribution
to $f^\lambda$ is
\begin{equation}
  f^\lambda=-\f{1}{R_\lambda}\p_R(\Phi_2\half z^2)=\f{3}{2}\f{GMz^2}{R^4R_\lambda}.
\end{equation}
Its contribution to $\ell\mathcal{M}(\p_t+\bar v^\lambda\p_\lambda)E$
is therefore
\begin{eqnarray}
  \lefteqn{\iint J\rho\left(-\rmi\lambda R_\lambda f^\lambda\,\rme^{\rmi\phi}\right)\rmd\phi\,\rmd z}&\nonumber\\
  &&=-\f{3\rmi}{2}\int\Omega^2\,\rme^{\rmi\phi}\left(\int J\rho z^2\,\rmd z\right)\rmd\phi,
\end{eqnarray}
which is the first term on the right-hand side of equation~(167) of
\citet{2001MNRAS.325..231O}.  This is a three-dimensional effect due
to the weakening of the (cylindrical) radial gravitational force away
from the midplane.  The cylindrical radial component is the relevant
one because the gas away from the midplane is moving in a plane of
(approximately) constant $z$, rather than in an inclined Keplerian
orbit.

Although the integrals of stresses and forces that appear in the
evolutionary equations involve integrals with respect to the azimuthal
angle and are in this sense global or large-scale quantities, we will
see below that these integrals naturally emerge in the form of
time-averages in a local model that follows the orbital motion.

\section{Local model of an eccentric disc}
\label{s:local}

\subsection{Flow decomposition}
\label{s:decomposition}

We have seen in Section~\ref{s:orbital} that the eccentric orbital
motion with $u^\phi=\Omega(\lambda,\phi)$ and $u^\lambda=u^z=0$
satisfies the equation of motion in the midplane $z=0$ when the
pressure is neglected.  We now write the fluid motion as the sum of
this orbital motion and a relative velocity~$v^i$:
\begin{equation}
  u^\lambda=v^\lambda,\qquad
  u^\phi=\Omega+v^\phi,\qquad
  u^z=v^z.
\end{equation}
The residual parts of the hydrodynamic equations
(\ref{dul})--(\ref{dupressure}) are then, without approximation,
\begin{eqnarray}
  \lefteqn{\rmD v^\lambda+\Gamma^\lambda_{\lambda\lambda}(v^\lambda)^2+2\Gamma^\lambda_{\lambda\phi}v^\lambda(\Omega+v^\phi)+\Gamma^\lambda_{\phi\phi}(2\Omega+v^\phi)v^\phi}&\nonumber\\
  &&=-g^{\lambda\lambda}\left[\p_\lambda(\Phi-\Phi_0)+\f{1}{\rho}\p_\lambda p\right]\nonumber\\
  &&\quad-g^{\lambda\phi}\left[\p_\phi(\Phi-\Phi_0)+\f{1}{\rho}\p_\phi p\right],
\end{eqnarray}
\begin{eqnarray}
  \lefteqn{\rmD v^\phi+(v^\lambda\p_\lambda+v^\phi\p_\phi)\Omega+2\Gamma^\phi_{\lambda\phi}v^\lambda(\Omega+v^\phi)+\Gamma^\phi_{\phi\phi}(2\Omega+v^\phi)v^\phi}&\nonumber\\
   &&=-g^{\lambda\phi}\left[\p_\lambda(\Phi-\Phi_0)+\f{1}{\rho}\p_\lambda p\right]\nonumber\\
  &&\quad-g^{\phi\phi}\left[\p_\phi(\Phi-\Phi_0)+\f{1}{\rho}\p_\phi p\right],
\end{eqnarray}
\begin{equation}
  \rmD v^z=-\p_z(\Phi-\Phi_0)-\f{1}{\rho}\p_zp,
\end{equation}
\begin{equation}
  \rmD\rho=-\rho\left[\Delta+\f{1}{J}\p_\lambda(Jv^\lambda)+\f{1}{J}\p_\phi(Jv^\phi)+\p_zv^z\right],
\end{equation}
\begin{equation}
  \rmD p=-\gamma p\left[\Delta+\f{1}{J}\p_\lambda(Jv^\lambda)+\f{1}{J}\p_\phi(Jv^\phi)+\p_zv^z\right],
\end{equation}
where
\begin{equation}
  \rmD=\p_t+v^\lambda\p_\lambda+(\Omega+v^\phi)\p_\phi+v^z\p_z
\end{equation}
is the Lagrangian derivative and
\begin{equation}
  \Delta=\f{1}{J}\p_\phi(J\Omega)
\end{equation}
is the orbital velocity divergence, which vanishes only in the case
$E=\cst$.

To simplify the equations in a way appropriate for a local model of a
thin disc, we apply the following scaling argument.  Let
$\epsilon\ll1$ be a characteristic value of the aspect ratio $H/r$ of
the disc, and let us consider a system of units in which $r$ and
$\Omega$ are $O(\epsilon^0)$, so that $H$ and the sound speed are
$O(\epsilon^1)$.  Therefore $p/\rho$ and $\Phi-\Phi_0$ are
$O(\epsilon^2)$.  We are interested in describing local nonlinear
fluid dynamical phenomena that take place on a lengthscale comparable
to $H$ and on a timescale comparable to $\Omega^{-1}$.  Therefore,
when acting on $v^i$, $\rho$ or $p$, the operator $\p_i$ is
$O(\epsilon^{-1})$, while the operator $\rmD$ is $O(\epsilon^0)$.  We
assume that the relative velocity components $v^i$ are
$O(\epsilon^1)$, i.e.\ comparable in magnitude to the sound speed but
much smaller than the orbital velocity.  This allows them to be
nonlinear and to generate Reynolds stresses comparable to the
pressure.  The equations simplify at the leading order in $\epsilon$
to
\begin{equation}
  \rmD v^\lambda+2\Gamma^\lambda_{\lambda\phi}\Omega v^\lambda+2\Gamma^\lambda_{\phi\phi}\Omega v^\phi=-\f{1}{\rho}\left(g^{\lambda\lambda}\p_\lambda p+g^{\lambda\phi}\p_\phi p\right),
\label{dvl}
\end{equation}
\begin{eqnarray}
  \lefteqn{\rmD v^\phi+(v^\lambda\p_\lambda+v^\phi\p_\phi)\Omega+2\Gamma^\phi_{\lambda\phi}\Omega v^\lambda+2\Gamma^\phi_{\phi\phi}\Omega v^\phi}&\nonumber\\
  &&=-\f{1}{\rho}\left(g^{\lambda\phi}\p_\lambda p+g^{\phi\phi}\p_\phi p\right),
\label{dvp}
\end{eqnarray}
\begin{equation}
  \rmD v^z=-\Phi_2z-\f{1}{\rho}\p_zp,
\label{dvz}
\end{equation}
\begin{equation}
  \rmD\rho=-\rho(\Delta+\p_\lambda v^\lambda+\p_\phi v^\phi+\p_zv^z),
\label{dvrho}
\end{equation}
\begin{equation}
  \rmD p=-\gamma p(\Delta+\p_\lambda v^\lambda+\p_\phi v^\phi+\p_zv^z),
\label{dvpressure}
\end{equation}
where $\Phi_2$ is defined in equation~(\ref{taylor}).  Although
several terms involving products of components of $v^i$ have been
dropped in equations (\ref{dvl}) and (\ref{dvp}), these equations are
still nonlinear because $v^i$ appears in the Lagrangian derivative.
The `inertial' terms involving the interaction between the orbital
motion and the relative velocity are linear, however.  The Jacobian
$J$ has dropped out of equations~(\ref{dvrho}) and~(\ref{dvpressure})
because it varies on a lengthscale $O(\epsilon^0)$.  For similar
reasons there are no horizontal derivatives of $\Phi_2$ in these
equations.

\subsection{Local approximation}

\begin{figure}
\subfigure{\includegraphics[trim=0cm 0cm 0cm 0cm, clip=true,width=0.44\textwidth]{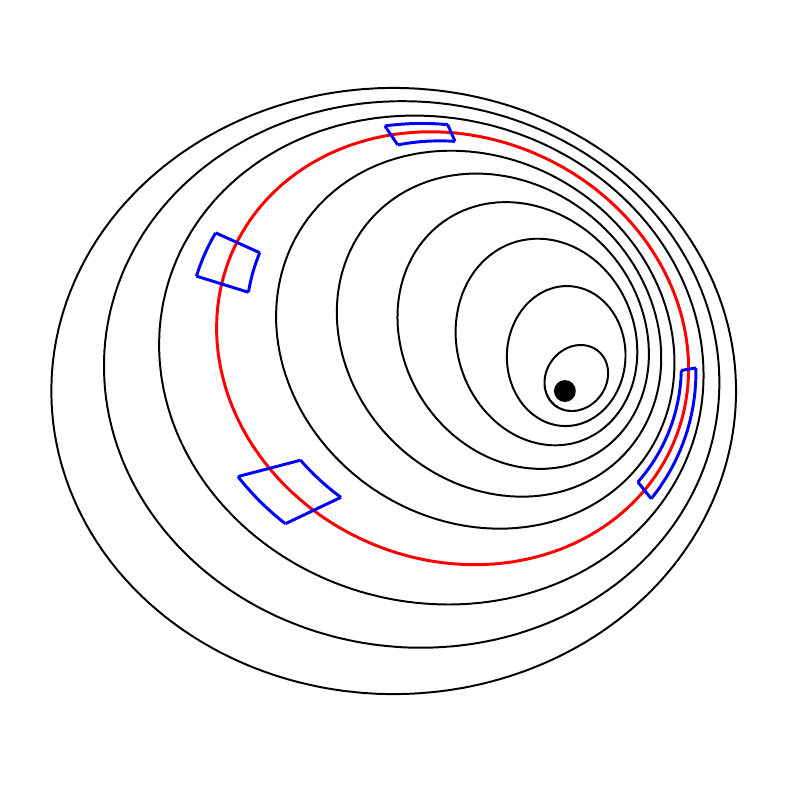} }
\caption{Construction of a local model of an eccentric disc.  The
  local orbital coordinates $(\xi,\eta,\zeta)$ are centred on a
  reference point that follows a particular Keplerian orbit (red
  ellipse).  The blue boxes represent the azimuthal compression due to
  the non-uniformity of the orbital motion, discussed in
  Section~\ref{s:shearing}, although they do not represent the orbital
  shear.}
\label{f:local}
\end{figure}

We now select a reference orbit $\lambda=\lambda_0$ with angular
velocity $\Omega(\lambda_0,\phi)$ (Fig.~\ref{f:local}).  We consider a
reference point that follows this orbit, starting from the pericentre
$\phi=\omega_0$ at $t=0$.  Let $\varphi(t)$ be the solution of
$\rmd\varphi(t)/\rmd t=\Omega(\lambda_0,\varphi(t))$ subject to the
initial condition $\varphi(0)=\omega_0$.  Then the orbital coordinates
of the reference point are
$(\lambda,\phi,z)=(\lambda_0,\varphi(t),0)$.  Let $\mathcal{P}_0$ be
the period of the reference orbit, such that
$\varphi(\mathcal{P}_0)=\varphi(0)=\omega_0$.

We examine the neighbourhood of the reference point by letting
\begin{equation}
  \lambda=\lambda_0+\xi,\qquad
  \phi=\varphi(t)+\eta,\qquad
  z=\zeta,\qquad
  t=\tau,
\end{equation}
where $(\xi,\eta)$ are (non-orthogonal) local coordinates in the
orbital plane, while $\zeta$ and $\tau$ are introduced only for
notational uniformity.  As we are interested in a region with an
extent comparable to $H$ in each dimension around the reference point,
$\xi$, $\eta$ and $\zeta$ are small, $O(\epsilon)$.

Since we are evaluating equations (\ref{dvl})--(\ref{dvrho}) at the
leading order in $\epsilon$, the geometrical coefficients appearing in
those equations may be replaced with their values at the reference
point $(\lambda,\phi)=(\lambda_0,\varphi(\tau))$, making them
functions of $\tau$ only, with period $\mathcal{P}_0$.

Because of its appearance in the Lagrangian derivative, however, the
orbital angular velocity $\Omega(\lambda,\phi)$ needs to be expanded
about the reference point in the form
\begin{equation}
  \Omega=\Omega_0(\tau)+\Omega_{\lambda0}(\tau)\xi+\Omega_{\phi0}(\tau)\eta+O(\epsilon^2),
\end{equation}
where $\Omega_0(\tau)=\Omega(\lambda_0,\varphi(\tau))$ is the angular
velocity of the reference point, while $\Omega_{\lambda0}(\tau)$ and
$\Omega_{\phi0}(\tau)$ are the partial derivatives $\p_\lambda\Omega$
and $\p_\phi\Omega$ of $\Omega(\lambda,\phi)$ evaluated at that point.

Since, at a fixed time, the differentials of the local coordinates are
identical to those of the global orbital coordinates, the
contravariant velocity components and spatial derivatives are
unchanged by the transformation.  Thus
$(v^\xi,v^\eta,v^\zeta)=(v^\lambda,v^\phi,v^z)$ and
$(\p_\xi,\p_\eta,\p_\zeta)=(\p_\lambda,\p_\phi,\p_z)$.
However, the time-dependence of the transformation means that the
time-derivatives are related by
\begin{equation}
  \p_t+\Omega_0\p_\phi=\p_\tau.
\end{equation}
At the leading order in $\epsilon$, therefore,
\begin{equation}
  \rmD=\p_\tau+v^\xi\p_\xi+(\Omega_{\lambda0}\xi+\Omega_{\phi0}\eta+v^\eta)\p_\eta+v^\zeta\p_\zeta.
\end{equation}

Equations (\ref{dvl})--(\ref{dvrho}) are then rewritten as
\begin{equation} 
  \rmD v^\xi+2\Gamma^\lambda_{\lambda\phi}\Omega v^\xi+2\Gamma^\lambda_{\phi\phi}\Omega v^\eta=-\f{1}{\rho}\left(g^{\lambda\lambda}\p_\xi p+g^{\lambda\phi}\p_\eta p\right),
\label{dvxi}
\end{equation}
\begin{eqnarray}
  \lefteqn{\rmD v^\eta+(\Omega_\lambda+2\Gamma^\phi_{\lambda\phi}\Omega)v^\xi+(\Omega_\phi+2\Gamma^\phi_{\phi\phi}\Omega)v^\eta}&\nonumber\\
  &&=-\f{1}{\rho}\left(g^{\lambda\phi}\p_\xi p+g^{\phi\phi}\p_\eta p\right),
\label{dveta}
\end{eqnarray}
\begin{equation}
  \rmD v^\zeta=-\Phi_2\zeta-\f{1}{\rho}\p_\zeta p,
\label{dvzeta}
\end{equation}
\begin{equation}
  \rmD\rho=-\rho(\Delta+\p_\xi v^\xi+\p_\eta v^\eta+\p_\zeta v^\zeta),
\label{drho_local}
\end{equation}
in which, as explained above, the geometrical coefficients are
evaluated at the reference point
$(\lambda,\phi,z)=(\lambda_0,\varphi(\tau),0)$, which makes them
periodic functions of $\tau$.  We have also dropped the subscript
zeros, so that
\begin{equation}
  \rmD=\p_\tau+v^\xi\p_\xi+(\Omega_\lambda\xi+\Omega_\phi\eta+v^\eta)\p_\eta+v^\zeta\p_\zeta.
\end{equation}
The thermal energy equation can be written as either
\begin{equation}
  \rmD s=0
\end{equation}
or
\begin{equation}
  \rmD p=-\gamma p(\Delta+\p_\xi v^\xi+\p_\eta v^\eta+\p_\zeta v^\zeta).
\end{equation}
These are the equations of ideal hydrodynamics in the local model of
an eccentric disc in non-shearing coordinates.  In
Appendix~\ref{s:coefficients} we give expressions for the coefficients
appearing in these equations.  Since these involve the true anomaly
$\theta$ explicitly, and it is not straightforward to express $\theta$
in terms of the time $\tau$, it may be preferable to regard $\theta$
as the timelike variable instead of $\tau$, replacing the derivative
$\p_\tau$ with $\Omega\p_\theta$.

The metric coefficients are now functions of $\tau$ satisfying $\dot
g_{ij}=\Omega(g_{ik}\Gamma^k_{j\phi}+g_{kj}\Gamma^k_{i\phi})$, where
the dot denotes a time-derivative.  Similarly the Jacobian and the
orbital velocity divergence become functions of $\tau$ related by
$\Delta=(\dot J/J)+\Omega_\phi$, with $\Omega_\phi=\dot\Omega/\Omega$.
Thus the conservative form of equation (\ref{drho_local}) is
\begin{equation}
  \p_\tau(J\rho)+\p_\xi(J\rho v^\xi)+\p_\eta[J\rho(\Omega_\lambda\xi+\Omega_\phi\eta+v^\eta)]+\p_\zeta(J\rho v^\zeta)=0.
\end{equation}

The Lagrangian derivative of the specific kinetic energy of the
relative motion can be shown to be
\begin{equation}
  \rmD(\half g_{ij}v^iv^j)=-v^iv_j\nabla_i\omega^j-\Phi_2\zeta v^\zeta-\f{1}{\rho}v^i\p_ip.
\end{equation}
The first source term on the right-hand side involves the covariant
derivative $\nabla_i\omega^j=\p_i\omega^j+\Gamma^j_{ik}\omega^k$ of
the orbital velocity field.  Here $v_i=g_{ij}v^j$ are the covariant
components of the relative velocity.  The conservative form of the
energy equation for the relative motion is
\begin{eqnarray}
  \lefteqn{\p_\tau(J\rho\mathcal{E}_\mathrm{rel})+\p_\xi[J(\rho\mathcal{E}_\mathrm{rel}+p)v^\xi]}&\nonumber\\
  &&+\p_\eta[J\rho\mathcal{E}_\mathrm{rel}(\Omega_\lambda\xi+\Omega_\phi\eta+v^\eta)+Jpv^\eta]\nonumber\\
  &&+\p_\zeta[J(\rho\mathcal{E}_\mathrm{rel}+p)v^\zeta]\nonumber\\
  &&=J\rho(-v^iv_j\nabla_i\omega^j+\dot\Phi_2\half\zeta^2)-Jp\Delta,
\end{eqnarray}
where
\begin{equation}
  \mathcal{E}_\mathrm{rel}=\half g_{ij}v^iv^j+\varepsilon+\half\Phi_2\zeta^2.
\end{equation}
Note that the possible sources of energy for the local model are the
orbital shear (accessed through Reynolds stresses), the
time-dependence of the vertical gravity coefficient $\Phi_2$, and the
orbital velocity divergence.

\subsection{Shearing coordinates}
\label{s:shearing}

The equations of the local model derived so far have an explicit
dependence on the horizontal coordinates $\xi$ and $\eta$ because of
their appearance in the Lagrangian derivative (but only for
`non-axisymmetric' solutions that depend on~$\eta$).

We can derive a horizontally homogeneous model by transforming to
shearing (and oscillating) local coordinates
$(\xi',\eta',\zeta',\tau')$ that are Lagrangian with respect to the
local orbital motion and are defined by
\begin{equation}
  \xi'=\xi,\qquad
  \eta'=\alpha(\tau)\eta+\beta(\tau)\xi,\qquad
  \zeta'=\zeta,\qquad
  \tau'=\tau,
\end{equation}
where $\alpha$ and $\beta$ remain to be determined.  The scale factor
$\alpha$ corresponds to a time-dependent stretching of the azimuthal
coordinate (cf.\ Fig.~\ref{f:local}), while the term $\beta$
corresponds to a shearing of the coordinate system.  Partial
derivatives transform according to
\begin{equation}
  \p_\xi=\p_\xi'+\beta\p_\eta',\qquad
  \p_\eta=\alpha\p_\eta',\qquad
  \p_\zeta=\p_\zeta',
\end{equation}
\begin{equation}
  \p_\tau=\p_\tau'+(\dot\alpha\eta+\dot\beta\xi)\p_\eta'.
\end{equation}
The Jacobian of the shearing coordinates is $\mathcal{J}=J/\alpha$,
the factor of $\alpha$ coming from the transformation just introduced.
We will \textit{not} transform the velocity components because we are
interested in evaluating Reynolds stress coefficients such as
$T^{\lambda\phi}=T^{\xi\eta}$.  The Lagrangian derivative transforms
into the spatially homogeneous form
\begin{equation}
  \rmD=\p_\tau'+v^\xi(\p_\xi'+\beta\p_\eta')+\alpha v^\eta\p_\eta'+v^\zeta\p_\zeta'
\end{equation}
provided that $\alpha$ and $\beta$ are chosen such that
\begin{equation}
  \dot\alpha=-\alpha\Omega_\phi,\qquad
  \dot\beta=-\alpha\Omega_\lambda.
\end{equation}
Since $\Omega_\phi=\dot\Omega/\Omega$, the first equation is satisfied
by choosing $\alpha\propto\Omega^{-1}$.  This periodically variable
scale factor for the azimuthal direction means that the new coordinate
$\eta'$ is related to the mean anomaly or mean longitude, i.e.\ to the
time taken to travel along the orbit.  We choose $\alpha$ to have
dimensions of length, so that the new variable $\eta'$ does also.  The
factor $\beta$ has a secular dependence on time, representing orbital
shear, as well as a periodic one.  The expressions for $\alpha$ and
$\beta$ are
\begin{equation}
  \alpha=\lambda(1+e\cos\theta)^{-2},
\end{equation}
\begin{eqnarray}
  \lefteqn{\beta=\f{3}{2}\left(1+\f{2e\lambda e'}{1-e^2}\right)\left(\f{GM}{\lambda^3}\right)^{1/2}\tau}&\nonumber\\
  &&-\f{\lambda e'(2+e\cos\theta)\sin\theta}{(1-e^2)(1+e\cos\theta)^2}-\f{\lambda\omega'}{(1+e\cos\theta)^2}\nonumber\\
  &&+\cst.
\label{beta}
\end{eqnarray}
The relation between the time $\tau$ (with origin $\tau=0$ at
pericentre) and $\theta$ is given by Kepler's equation in the form
\begin{equation}
  n\tau=\mathcal{E}-e\sin\mathcal{E},
\end{equation}
where $n=(GM/a^3)^{1/2}$ is the mean motion, $a=\lambda/(1-e^2)$ is the
semi-major axis and $\mathcal{E}$ is the eccentric anomaly, such that
\begin{equation}
  \sin\mathcal{E}=\f{(1-e^2)^{1/2}\sin\theta}{1+e\cos\theta}.
\end{equation}
Thus $\tau$ can be expressed in terms of $\theta$ by inverting this
sine function.  The explicit expression for the Jacobian of the
shearing coordinates is
\begin{equation}
  \mathcal{J}=\f{1+(e-\lambda e')\cos\theta-\lambda e\omega'\sin\theta}{1+e\cos\theta}.
\end{equation}

The equations of the local model now have the form
\begin{eqnarray} 
  \lefteqn{\rmD v^\xi+2\Gamma^\lambda_{\lambda\phi}\Omega v^\xi+2\Gamma^\lambda_{\phi\phi}\Omega v^\eta}&\nonumber\\
  &&=-\f{1}{\rho}\left[g^{\lambda\lambda}(\p_\xi'+\beta\p_\eta')p+g^{\lambda\phi}\alpha\p_\eta'p\right],
\end{eqnarray}
\begin{eqnarray}
  \lefteqn{\rmD v^\eta+(\Omega_\lambda+2\Gamma^\phi_{\lambda\phi}\Omega)v^\xi+(\Omega_\phi+2\Gamma^\phi_{\phi\phi}\Omega)v^\eta}&\nonumber\\
  &&=-\f{1}{\rho}\left[g^{\lambda\phi}(\p_\xi'+\beta\p_\eta')p+g^{\phi\phi}\alpha\p_\eta'p\right],
\end{eqnarray}
\begin{equation}
  \rmD v^\zeta=-\Phi_2\zeta'-\f{1}{\rho}\p_\zeta'p,
\end{equation}
\begin{equation}
  \rmD\rho=-\rho\left[\Delta+(\p_\xi'+\beta\p_\eta')v^\xi+\alpha\p_\eta'v^\eta+\p_\zeta'v^\zeta\right].
\end{equation}
The thermal energy equation can be written as either
\begin{equation}
  \rmD s=0
\end{equation}
or
\begin{equation}
  \rmD p=-\gamma p\left[\Delta+(\p_\xi'+\beta\p_\eta')v^\xi+\alpha\p_\eta'v^\eta+\p_\zeta'v^\zeta\right].
\end{equation}
In Appendix~\ref{s:mhd} we give, for future reference, the extension
of these equations to ideal magnetohydrodynamics.

Significantly, the coefficients in these equations are independent of
$\xi'$ and $\eta'$: the model is homogeneous in the local horizontal
spatial coordinates.  However, the coefficients do depend on $\tau'$.
This time-dependence is mostly periodic, with the orbital period, but
terms associated with azimuthal derivatives ($\p_\eta'$) have in
addition a secular dependence on $\tau'$ through $\beta$.

This spatial homogeneity means that periodic boundary conditions can
be applied in $\xi'$ and $\eta'$, leading to the model of an eccentric
shearing sheet or box.  In this case $\xi'$ and $\eta'$ are restricted
to the ranges $[-L_{\xi'}/2,L_{\xi'}/2]$ and
$[-L_{\eta'}/2,L_{\eta'}/2]$ respectively, identifying a finite patch
of fluid.  However, because of the secular dependence of $\beta$ on
$\tau'$, the shearing coordinates should be remapped from time to time
to prevent them from becoming too distorted.  This can be done, for
example, by reducing the value of the additive constant in
equation~(\ref{beta}) from time to time so that $\beta$ remains within
a reasonable range of positive and negative values.  When $\beta$ is
changed from $\beta_1$ to $\beta_2$, the relationship between the
non-shearing coordinates $(\xi,\eta)$ and the shearing coordinates
$(\xi',\eta')$ is modified; in order for the periodic boundary
conditions to be mutually compatible in the old and new coordinates,
it is necessary that $\beta_1-\beta_2$ be an integer multiple of the
aspect ratio $L_\eta'/L_\xi'$.

The local model inherits three dimensionless parameters from the
geometry of the eccentric disc: $e$ (eccentricity), $\lambda e'$
(eccentricity gradient) and $\lambda e\omega'$ (twist).  This makes it
more complicated than the local model of warped discs
\citep{2013MNRAS.433.2403O}.

Using the fact that the orbital velocity divergence is
$\Delta=\p_\tau\ln(J\Omega)=\p_\tau\ln(J/\alpha)=\p_\tau\ln\mathcal{J}$,
we can express the conservation of mass in the form
\begin{equation}
  \p_\tau'(\mathcal{J}\rho)+\p_\xi'(\mathcal{J}\rho v^\xi)+\p_\eta'[\mathcal{J}\rho(\alpha v^\eta+\beta v^\xi)]+\p_\zeta'(\mathcal{J}\rho v^\zeta)=0.
\end{equation}
Subject to periodic boundary conditions in $\xi'$ and $\eta'$, and
suitable boundary conditions in $\zeta'$, the conserved mass in the
shearing box is
\begin{equation}
  \int\,\rmd M=\iiint\mathcal{J}\rho\,\rmd\xi'\,\rmd\eta'\,\rmd\zeta'.
\end{equation}

The horizontal momentum components $P^\xi=\int v^\xi\,\rmd M$ and $P^\eta=\int v^\eta\,\rmd M$ of the box satisfy the equations
\begin{equation}
  \p_\tau'P^\xi+2\Gamma^\lambda_{\lambda\phi}\Omega P^\xi+2\Gamma^\lambda_{\phi\phi}\Omega P^\eta=0,
\end{equation}
\begin{equation}
  \p_\tau'P^\eta+(\Omega_\lambda+2\Gamma^\phi_{\lambda\phi}\Omega)P^\xi+(\Omega_\phi+2\Gamma^\phi_{\phi\phi}\Omega)P^\eta=0,
\end{equation}
which allow an epicyclic oscillation around the reference orbit, but
it would be natural to constrain the solutions to satisfy
$P^\xi=P^\eta=0$, meaning that the reference orbit has been correctly
defined.

The energy equation in shearing coordinates has the form
\begin{eqnarray}
  \lefteqn{\p_\tau'(\mathcal{J}\rho\mathcal{E}_\mathrm{rel})+\p_\xi'[\mathcal{J}(\rho\mathcal{E}_\mathrm{rel}+p)v^\xi]}&\nonumber\\
  &&+\p_\eta'[\mathcal{J}(\rho\mathcal{E}_\mathrm{rel}+p)(\alpha v^\eta+\beta v^\xi)]\nonumber\\
  &&+\p_\zeta[\mathcal{J}(\rho\mathcal{E}_\mathrm{rel}+p)v^\zeta]\nonumber\\
  &&=\mathcal{J}\rho(-v^iv_j\nabla_i\omega^j+\dot\Phi_2\half\zeta^2)-\mathcal{J}p\Delta.
\end{eqnarray}

In Section~\ref{s:evolution} we saw that four different integrals of
components of the stress tensor, (\ref{stress1})--(\ref{stress4}), are
required in order to close the system of equations governing the
global evolution of the shape and mass distribution of an eccentric
disc.  In the local model the stress components can readily be
calculated; for example, the Reynolds-stress component
$T^{\lambda\phi}$ corresponds to $-\rho v^\xi v^\eta$.  The
geometrical factors $J$, $R$, $\rme^{\rmi\phi}$ and $R_\lambda$ are
known functions of time in the local model.  Since the local model
follows an orbiting reference point, the azimuthal integral
$\int\dots\,\rmd\phi$ can be interpreted as a time-integral
$\int\dots\Omega\,\rmd\tau'$ over a single orbit, where $\Omega$ is
also a known function of time.  In the case of turbulent flows,
additional spatial averaging over the box and time-averaging over
multiple orbits can be carried out to obtain the relevant stress
integrals.

\subsection{Relation to the standard shearing sheet}

If $e=0$, the reference orbit is circular and the model can be related
to the standard shearing sheet.  We then have $\lambda=R$ and
$\Omega=(GM/R^3)^{1/2}=\cst$ on the reference orbit, and the metric
components and connection coefficients simplify considerably.

The equations of the local model (in non-shearing coordinates) reduce
to
\begin{equation}
  \rmD v^\xi+2\Delta v^\xi-\f{2R\Omega}{\mathcal{J}}v^\eta=-\f{1}{\rho\mathcal{J}^2}\p_\xi p,
\end{equation}
\begin{equation}
  \rmD v^\eta+\f{\Omega}{2R}v^\xi=-\f{1}{\rho R^2}\p_\eta p,
\end{equation}
\begin{equation}
  \rmD v^\zeta=-\Omega^2\zeta-\f{1}{\rho}\p_\zeta p,
\end{equation}
\begin{equation}
  \rmD\rho=-\rho(\Delta+\p_\xi v^\xi+\p_\eta v^\eta+\p_\zeta v^\zeta),
\end{equation}
with
\begin{equation}
  \rmD=\p_\tau+v^\xi\p_\xi+\left[(\half-2\mathcal{J})\f{\Omega\xi}{R}+v^\eta\right]\p_\eta+v^\zeta\p_\zeta,
\end{equation}
\begin{equation}
  \Delta=\left(\f{\lambda e'\sin\theta}{1-\lambda e'\cos\theta}\right)\Omega=\f{\dot{\mathcal{J}}}{\mathcal{J}},
\end{equation}
\begin{equation}
  \mathcal{J}=\f{J}{R}=R_\lambda=1-\lambda e'\cos\theta
\end{equation}
and $\theta=\Omega t$.

These equations can be derived from the standard hydrodynamic
equations of the shearing sheet,
\begin{equation}
  \rmD v^x-2\Omega v^y=-\f{1}{\rho}\p_xp,
\end{equation}
\begin{equation}
  \rmD v^y+\f{1}{2}\Omega v^x=-\f{1}{\rho}\p_yp,
\end{equation}
\begin{equation}
  \rmD v^z=-\Omega^2z-\f{1}{\rho}\p_zp,
\end{equation}
\begin{equation}
  \rmD\rho=-\rho(\p_xv^x+\p_yv^y+\p_zv^z),
\end{equation}
with
\begin{equation}
  \rmD=\p_t+v^x\p_x+(-{\textstyle\f{3}{2}}\Omega x+v^y)\p_y+v^z\p_z.
\end{equation}
(Note that, since these are Cartesian coordinates, $v^x$ is the same
as $v_x$, etc.)

If $e'=0$ then the relation is straightforward because $\mathcal{J}=1$
and $\Delta=0$.  Note that $Rv^\eta$ equates to $v^y$, and
$(1/R)\p_\eta$ to $\p_y$, because $\eta$ is an angular variable.

If $e'\ne0$ then a time-dependent homogeneous transformation of the
radial coordinate is involved
\citep[cf.][Appendix~A]{2009Icar..202..565L}:
\begin{equation}
  x=\mathcal{J}\xi,\qquad
  t=\tau,
\end{equation}
so that
\begin{equation}
  \p_x=\mathcal{J}^{-1}\p_\xi,\qquad
  \p_t=\p_\tau-\Delta\xi\p_\xi,
\end{equation}
and the velocity components are related by
\begin{equation}
  v^x=\mathcal{J}v^\xi+\dot{\mathcal{J}}\xi,\qquad
  v^y=Rv^\eta+\half(1-\mathcal{J})\Omega\xi.
\end{equation}
The solution with $v^\xi=v^\eta=0$ corresponds to a free epicyclic
oscillation proportional to $\xi$, which is the local representation
of the eccentricity gradient.

Models of this type have been used in the theory of planetary rings
\citep{1996Icar..122..128M}, in which the parameter $|\lambda e'|<1$
is called~$q$.

\section{Laminar flows}
\label{s:laminar}

\subsection{Nonlinear vertical oscillations}

Laminar flows are the simplest solutions of the local model, being
horizontally invariant and having a purely vertical velocity $v^\zeta$
that, like $\rho$ and $p$, depends only on $\zeta$ and $\tau$.  They
satisfy the equations
\begin{equation}
  \rmD v^\zeta=-\Phi_2\zeta-\f{1}{\rho}\p_\zeta p,
\label{dvzlaminar}
\end{equation}
\begin{equation}
  \rmD\rho=-\rho(\Delta+\p_\zeta v^\zeta),
\end{equation}
\begin{equation}
  \rmD p=-\gamma p(\Delta+\p_\zeta v^\zeta),
\label{dvplaminar}
\end{equation}
with $\rmD=\p_\tau+v^\zeta\p_\zeta$.

The periodic variation of the vertical gravity coefficient $\Phi_2$
around the orbit (in the case $E\ne0$) and the orbital velocity
divergence $\Delta$ (in the case $E'\ne0$) drive a vertical
oscillation of the disc, which is nonlinear unless $E$ and $\lambda
E'$ are both small.

Before solving these equations we consider the simpler problem of
hydrostatic equilibrium in a circular disc where the vertical
gravitational acceleration is proportional to the distance above the
midplane.  Let $F_\rho(x)$ and $F_p(x)$ be dimensionless functions of
a dimensionless vertical coordinate $x$, which describe the
equilibrium profiles of density and pressure.  The equation of
hydrostatic equilibrium in dimensionless form is
\begin{equation}
  F_p'(x)=-xF_\rho(x).
\end{equation}
We normalize the profiles such that
\begin{equation}
  \int_{-\infty}^\infty F_\rho(x)\,\rmd x=1
\end{equation}
and
\begin{equation}
  \int_{-\infty}^\infty F_p(x)\,\rmd x=\int_{-\infty}^\infty F_\rho(x)x^2\,\rmd x=1.
\end{equation}
(The latter two integrals are easily shown to be equal by integrating
by parts and applying reasonable boundary conditions.)  Simple
examples are the isothermal structure,
\begin{equation}
  F_\rho(x)=F_p(x)=(2\pi)^{-1/2}\exp\left(-\f{x^2}{2}\right),
\end{equation}
the homogeneous structure,
\begin{equation}
  F_\rho=\f{1}{2\sqrt{3}},
\end{equation}
\begin{equation}
  F_p=\f{3-x^2}{4\sqrt{3}}
\end{equation}
(for $x^2<3$ only), and the polytropic structure,
\begin{equation}
  F_\rho(x)=C_n\left(1-\f{x^2}{2n+3}\right)^n,
\end{equation}
\begin{equation}
  F_p(x)=\f{2n+3}{2(n+1)}C_n\left(1-\f{x^2}{2n+3}\right)^{n+1},
\end{equation}
(for $x^2<2n+3$ only) where $n>0$ (not necessarily an integer) is the
polytropic index and
\[
  C_n=[(2n+3)\pi]^{-1/2}\f{\Gamma(n+{\textstyle\f{3}{2}})}{\Gamma(n+1)}
\]
is a normalization constant.  It can be shown that the polytropic
structure approaches the isothermal structure in the limit
$n\to\infty$, and approaches the homogeneous structure in the limit
$n\to0$.

Provided that $\gamma=\cst$, the laminar flow has the form of a
homogeneous expansion and contraction of the disc,
\begin{equation}
  v^\zeta=w(\tau)\zeta,
\end{equation}
\begin{equation}
  \rho=\hat\rho(\tau)F_\rho(x),
\end{equation}
\begin{equation}
  p=\hat p(\tau)F_p(x),
\end{equation}
where $x=\zeta/H(\tau)$ is the vertical coordinate $\zeta$ scaled by a
time-dependent vertical scaleheight $H(\tau)$, and the dimensionless
functions $F_\rho(x)$ and $F_p(x)$ are as defined above.  Although the
disc is not in hydrostatic equilibrium, its internal structure can be
related to that of an equilibrium disc.  In the isothermal case
$H(\tau)$ is the Gaussian scaleheight of the disc, while in the
homogeneous case or the polytropic case it is a fraction of the true
semi-thickness. In each case the surface density is
$\Sigma(\tau)=\int\rho\,\rmd\zeta=\hat\rho(\tau)H(\tau)$ and the
second vertical moment of the density is
$\int\rho\zeta^2\,\rmd\zeta=\hat\rho(\tau)H(\tau)^3$, so $H(\tau)$ is
the standard deviation of the mass distribution.

Equations (\ref{dvzlaminar})--(\ref{dvplaminar}) are satisfied provided that the
functions $H(\tau)$, $w(\tau)$, $\hat\rho(\tau)$ and $\hat p(\tau)$ obey
\begin{equation}
  \f{\dot H}{H}=w,
\label{hdot}
\end{equation}
\begin{equation}
  \dot w+w^2=-\Phi_2+\f{\hat p}{\hat\rho H^2},
\label{wdot}
\end{equation}
\begin{equation}
  \f{\dot{\hat\rho}}{\hat\rho}=\f{\dot{\hat p}}{\gamma\hat p}=-(\Delta+w).
\label{rhohatdot}
\end{equation}

Note that, if $\gamma>1+n^{-1}$ (or $\gamma>1$ in the case of an
isothermal structure), the disc is stably stratified.  However,
buoyancy forces do not affect the dynamics of the laminar flow because
of its horizontal invariance.

The surface density satisfies
\begin{equation}
  \f{\dot\Sigma}{\Sigma}=\f{\dot{\hat\rho}}{\hat\rho}+\f{\dot H}{H}=-\Delta=-\f{\dot J}{J}-\f{\dot\Omega}{\Omega}.
\end{equation}
This is a statement of mass conservation, and implies that
$J\Sigma\Omega=\cst$ \citep[cf.][]{2001MNRAS.325..231O}.  If the
orbital velocity divergence is non-zero because of an eccentricity
gradient, then the surface density varies periodically around the
orbit.  In terms of the one-dimensional mass density introduced in
Section~\ref{s:evolution},
\begin{equation}
  \mathcal{M}=\int J\Sigma\,\rmd\phi=\int J\Sigma\Omega\,\rmd\tau=J\Sigma\Omega\int\rmd\tau,
\end{equation}
we have $J\Sigma\Omega=\mathcal{M}/\mathcal{P}$, where $\mathcal{P}$
is the orbital period.

Since $\hat\rho\propto(J\Omega H)^{-1}$ and
$\hat p\propto\hat\rho^\gamma\propto(J\Omega H)^{-\gamma}$, the last term
in equation (\ref{wdot}) is $\propto(J\Omega H)^{-(\gamma-1)}H^{-2}$.
When $w$ is eliminated between equations (\ref{hdot}) and (\ref{wdot})
we obtain
\begin{equation}
  \f{\ddot H}{H}+\Phi_2\propto(J\Omega)^{-(\gamma-1)}H^{-(\gamma+1)},
\end{equation}
which describes a nonlinear vertical oscillator forced by the
periodically varying orbital geometry.

We are mainly interested in solutions that have period $\mathcal{P}$
in $\tau$ (or period $2\pi$ in $\theta$), which are stationary on the
orbital timescale when viewed in a non-rotating frame of reference.
The general solution, however, includes a free oscillation that can
only be eliminated by an appropriate choice of initial condition, or
by including some dissipation.

In fact it is straightforward to include a bulk viscosity in the
description of laminar flows.  If the dynamic bulk viscosity is
parametrized as $\mu_\rmb=\alpha_\rmb p(GM/\lambda^3)^{-1/2}$, as in
\citet{2001MNRAS.325..231O}, and $\alpha_\rmb$ is independent of
$\zeta$, then equation~(\ref{wdot}) is modified to
\begin{equation}
  \dot w+w^2=-\Phi_2+\left[1-\alpha_\rmb\left(\f{GM}{\lambda^3}\right)^{-1/2}(\Delta+w)\right]\f{\hat p}{\hat\rho H^2}.
\end{equation}
We neglect the effects of viscous heating.
[\citet{2001MNRAS.325..231O} included shear and bulk viscosity
(allowing for a non-zero relaxation time), viscous heating and
radiative cooling.]

\subsection{Linear theory for small eccentricity}

\begin{figure*}
\subfigure{\includegraphics[trim=0cm 0cm 0cm 0cm, clip=true,width=0.44\textwidth]{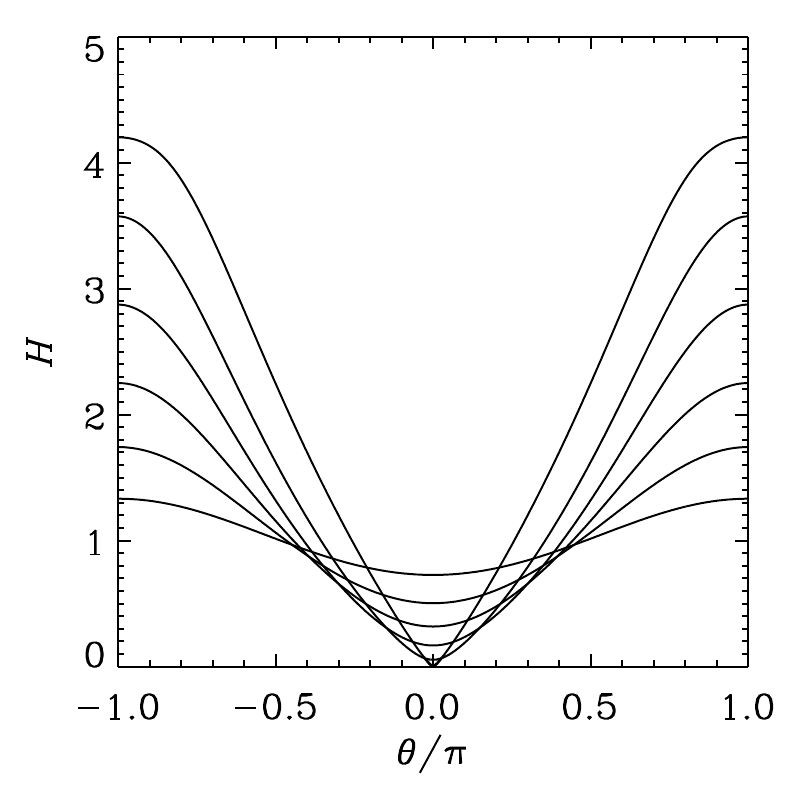} }
\subfigure{\includegraphics[trim=0cm 0cm 0cm 0cm, clip=true,width=0.44\textwidth]{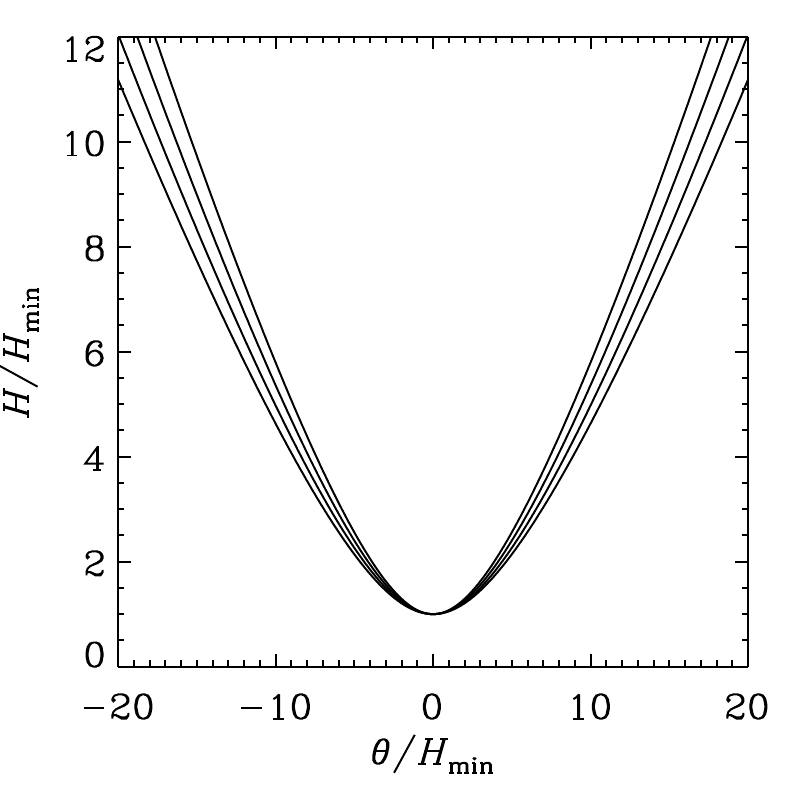} }
\caption{Left: Laminar flows for eccentric discs with eccentricities
  $0.1$, $0.2$, $0.3$, $0.4$, $0.5$ and $0.6$, no eccentricity
  gradient and $\gamma=1$.  The vertical scaleheight, in units of the
  hydrostatic value for a circular disc with the same semi-latus
  rectum, is plotted versus the true anomaly.  For small $e$ the
  variation is approximately sinusoidal and agrees with the linear
  theory, but for larger $e$ extreme behaviour occurs close to the
  pericentre where $H$ approaches zero.  Right: Behaviour near
  pericentre for $e=0.6$, $0.65$, $0.7$ and $0.75$ (inner to outer
  lines), with variables scaled by the corresponding minimum value of
  $H$, which is $H_\mathrm{min}=2.56\times10^{-3}$,
  $9.69\times10^{-5}$, $4.36\times10^{-7}$ and $3.05\times10^{-11}$,
  respectively.}
\label{f:h}
\end{figure*}

A linear theory can be developed when both $E$ and $\lambda E'$ are
small compared to unity, in which case the eccentric disc can be
regarded as a small perturbation of a circular disc in which $\lambda$
is the radial coordinate.  In the circular case, the local model is
hydrostatic with $H=H_0=\cst$, $w=0$, $\hat\rho=\hat\rho_0=\cst$ and
$\hat p=\hat p_0=\cst$, such that $\hat
p_0/\hat\rho_0H_0^2=\Phi_2=GM/\lambda^3$.

In the presence of a small eccentricity and a small eccentricity
gradient, such that $e$ and $\lambda e'$ are $O(\epsilon)$ with
$\epsilon\ll1$ being a small parameter (different from that used
previously), we have (from Appendix~\ref{s:coefficients})
\begin{eqnarray}
  \Phi_2&=&\left(\f{GM}{\lambda^3}\right)[1+3e\cos\theta+O(\epsilon^2)]\nonumber\\
  &=&\left(\f{GM}{\lambda^3}\right)\left[1+\mathrm{Re}\left(3E\,\rme^{-\rmi\phi}\right)+O(\epsilon^2)\right],
\end{eqnarray}
\begin{eqnarray}
  \Delta&=&\left(\f{GM}{\lambda^3}\right)^{1/2}[\lambda e'\sin\theta-\lambda e\omega'\cos\theta+O(\epsilon^2)]\nonumber\\
  &=&\left(\f{GM}{\lambda^3}\right)^{1/2}\left[\mathrm{Re}\left(\rmi\lambda E'\,\rme^{-\rmi\phi}\right)+O(\epsilon^2)\right].
\end{eqnarray}

The laminar solution is of the form
\begin{equation}
  H=H_0\left[1+\mathrm{Re}\left(\tilde H\,\rme^{-\rmi\phi}\right)+O(\epsilon^2)\right],
\end{equation}
\begin{equation}
  \hat\rho=\hat\rho_0\left[1+\mathrm{Re}\left(\tilde\rho\,\rme^{-\rmi\phi}\right)+O(\epsilon^2)\right],
\end{equation}
\begin{equation}
  \hat p=\hat p_0\left[1+\mathrm{Re}\left(\tilde p\,\rme^{-\rmi\phi}\right)+O(\epsilon^2)\right],
\end{equation}
\begin{equation}
  w=\left(\f{GM}{\lambda^3}\right)^{1/2}\left[\mathrm{Re}\left(\tilde w\,\rme^{-\rmi\phi}\right)+O(\epsilon^2)\right].
\end{equation}
Bearing in mind that $\p_\tau=\Omega\p_\theta$ with
$\Omega=(GM/\lambda^3)^{1/2}[1+O(\epsilon)]$, in order to satisfy
equations~(\ref{hdot})--(\ref{rhohatdot}) we require the dimensionless
perturbations to satisfy
\begin{equation}
  -\rmi\tilde H=\tilde w,
\end{equation}
\begin{equation}
  -\rmi\tilde w=-3E+\tilde p-\tilde\rho-2\tilde H-\alpha_\rmb(\rmi\lambda E'+\tilde w),
\label{itw}
\end{equation}
\begin{equation}
  -\rmi\tilde\rho=-\f{\rmi\tilde p}{\gamma}=-(\rmi\lambda E'+\tilde w),
\end{equation}
where we have allowed for a bulk viscosity as described above.  The
solution is
\begin{equation}
  \tilde H=\rmi\tilde w=\f{-3E+(\gamma-1-\rmi\alpha_\rmb)\lambda E'}{\gamma-\rmi\alpha_\rmb},
\label{htilde}
\end{equation}
\begin{equation}
  \tilde\rho=\f{\tilde p}{\gamma}=\f{3E+\lambda E'}{\gamma-\rmi\alpha_\rmb}.
\end{equation}
We can similarly define a dimensionless surface density perturbation
\begin{equation}
  \tilde\Sigma=\tilde\rho+\tilde H=\lambda E'.
\end{equation}
Note that this dynamical solution differs significantly from the
hydrostatic non-solution in which the acceleration ($-\rmi\tilde w$)
and viscous terms are neglected in equation~(\ref{itw}): $\tilde
H=\rmi\tilde w=[-3E+(\gamma-1)\lambda E']/(\gamma+1)$ and
$\tilde\rho=\tilde p/\gamma=(3E+2\lambda E')/(\gamma+1)$.
In the absence of bulk viscosity, the amplitude with which $H$
oscillates is larger by a factor of $1+\gamma^{-1}$ in the dynamical
solution than in the hydrostatic non-solution.

We now refer to the global analysis of Section~\ref{s:evolution} and
calculate the evolution of eccentricity associated with the laminar
solution.  The stress tensor of the laminar flow is
\begin{equation}
  T^{ij}=-p\left[1-\alpha_\rmb\left(\f{GM}{\lambda^3}\right)^{-1/2}(\Delta+w)\right]g^{ij}
\end{equation}
and its vertical integral is
\begin{equation}
  \int T^{ij}\,\rmd z=-\hat pH\left[1-\alpha_\rmb\left(\f{GM}{\lambda^3}\right)^{-1/2}(\Delta+w)\right]g^{ij}.
\end{equation}
In the linear theory developed above this becomes
\begin{equation}
  \int T^{\lambda\lambda}\,\rmd z=-\hat p_0H_0\left[1+\mathrm{Re}\left(\tilde T^{\lambda\lambda}\,\rme^{-\rmi\phi}\right)+O(\epsilon^2)\right]
\end{equation}
\begin{equation}
  \int\lambda T^{\lambda\phi}\,\rmd z=-\hat p_0H_0\left[\mathrm{Re}\left(\tilde T^{\lambda\phi}\,\rme^{-\rmi\phi}\right)+O(\epsilon^2)\right],
\end{equation}
\begin{equation}
  \int\lambda^2T^{\phi\phi}\,\rmd z=-\hat p_0H_0\left[1+\mathrm{Re}\left(\tilde T^{\phi\phi}\,\rme^{-\rmi\phi}\right)+O(\epsilon^2)\right],
\end{equation}
with
\begin{equation}
  \tilde T^{\lambda\lambda}=\tilde p+\tilde H-\alpha_\rmb(\rmi\lambda E'+\tilde w)+2(E+\lambda E'),
\end{equation}
\begin{equation}
  \tilde T^{\lambda\phi}=-\rmi E,
\end{equation}
\begin{equation}
  \tilde T^{\phi\phi}=\tilde p+\tilde H-\alpha_\rmb(\rmi\lambda E'+\tilde w)+2E,
\end{equation}
where the final terms in each case are due to the azimuthal variation
of the metric coefficients (Appendix~\ref{s:coefficients}).
The required stress integrals are therefore, correct to $O(\epsilon)$,
\begin{equation}
  \iint JR^2T^{\lambda\phi}\,\rmd\phi\,\rmd z=0,
\end{equation}
\begin{equation}
  \iint JR^2T^{\lambda\phi}\rme^{\rmi\phi}\,\rmd\phi\,\rmd z=-\pi\lambda^2\hat p_0H_0\tilde T^{\lambda\phi},
\end{equation}
\begin{equation}
  \iint JR_\lambda T^{\lambda\lambda}\rme^{\rmi\phi}\,\rmd\phi\,\rmd z=-\pi\lambda\hat p_0H_0\left(\tilde T^{\lambda\lambda}-3E-2\lambda E'\right),
\end{equation}
\begin{equation}
  \iint JR^2T^{\phi\phi}\rme^{\rmi\phi}\,\rmd\phi\,\rmd z=-\pi\lambda\hat p_0H_0\left(\tilde T^{\phi\phi}-4E-\lambda E'\right).
\end{equation}
To the same level of approximation,
\begin{equation}
  \mathcal{M}=\iint J\rho\,\rmd\phi\,\rmd z=2\pi\lambda\hat\rho_0H_0+O(\epsilon^2),
\end{equation}
\begin{equation}
  \iint J\rho\Omega^2z^2\,\rme^{\rmi\phi}\,\rmd\phi\,\rmd z=\pi\lambda\hat p_0H_0(2\tilde H+2E).
\end{equation}
We can now apply these results to the evolutionary
equation~(\ref{edot}) for $E$, evaluating it correct to first order.
Note that there is no angular momentum transport or quasi-radial mass
flux to this order because of the absence of shear viscosity.  In
applying our local results to the global disc, we write $\Sigma$ and
$P$ for vertically integrated density and pressure in the circular
global disc, and allow for the $\lambda$-dependence of these and other
quantities.  We obtain
\begin{eqnarray}
  \lefteqn{2\Sigma(GM\lambda^3)^{1/2}\p_tE=-\p_\lambda(2\lambda^2P\tilde T^{\lambda\phi})}&\nonumber\\
  &&+\rmi\lambda\p_\lambda\left[\lambda P(\tilde T^{\lambda\lambda}-3E-2\lambda E')\right]+\rmi\lambda P(\tilde T^{\phi\phi}-4E-\lambda E')\nonumber\\
  &&+\lambda P\tilde T^{\lambda\phi}-3\rmi\lambda P(\tilde H+E).
\end{eqnarray}
Using the above results this simplifies to
\begin{eqnarray}
  \lefteqn{2\Sigma(GM\lambda^3)^{1/2}\p_tE=\rmi\,\p_\lambda\left[\lambda^2P(\tilde H+4E+\lambda E')\right]}&\nonumber\\
  &&-\rmi\lambda P(3\tilde H+5E+\lambda E').
\end{eqnarray}
As can be seen from the expression~(\ref{htilde}) for $\tilde H$, this
equation contains $\gamma$ and $\alpha_\rmb$ only in the combination
$\gamma-\rmi\alpha_\rmb$ (as was also found in the two-dimensional
linear theory of \citealt{2006MNRAS.368.1123G}).  In the inviscid case
$\alpha_\rmb=0$ it simplifies to
\begin{eqnarray}
  \lefteqn{2\Sigma(GM\lambda^3)^{1/2}\p_tE=\rmi\,\p_\lambda[(2-\gamma^{-1})P\lambda^3E']}\nonumber\\
  &&+\rmi(4-3\gamma^{-1})\lambda^2E\p_\lambda P+3\rmi(1+\gamma^{-1})\lambda PE,
\label{edot3d}
\end{eqnarray}
which we have derived independently by a three-dimensional linear
perturbation analysis of a circular disc.  For comparison, the
two-dimensional linear theory of \citet{2006MNRAS.368.1123G} gives
instead
\begin{equation}
  2\Sigma(GM\lambda^3)^{1/2}\p_tE=\rmi\,\p_\lambda(\gamma P\lambda^3E')+\rmi\lambda^2E\p_\lambda P.
\end{equation}
Differences between the two- and three-dimensional theories were found
to be crucial for the dynamics of eccentric discs around Be stars by
\citet{2008MNRAS.388.1372O}.

Equation~(\ref{edot3d}) is a dispersive wave equation related to the
Schr\"odinger equation, and indicates how eccentricity propagates
through a disc by means of pressure.  Bulk viscosity can easily be
included by replacing $\gamma$ with $\gamma-\rmi\alpha_\rmb$; this
gives an eccentricity diffusion coefficient of
\begin{equation}
  \f{1}{2}\left(\f{\alpha_\rmb}{\gamma^2+\alpha_\rmb^2}\right)H^2\left(\f{GM}{\lambda^3}\right)^{1/2},
\end{equation}
which, for $\alpha_\rmb\ll\gamma$, is half the (mass-weighted mean)
kinematic bulk viscosity.

\subsection{Behaviour for larger eccentricity}

The linear theory shows that the periodic variation around the orbit
of the vertical gravity coefficient $\Phi_2$ and the orbital velocity
divergence $\Delta$ induce a dynamical vertical oscillation of the
disc.  For a disc that behaves isothermally ($\gamma=1$) the
fractional oscillation amplitude of the scaleheight in linear theory
is three times the eccentricity.  It is clear, then, that in this case
the oscillation will become strongly nonlinear at eccentricities well
below unity.

We have computed the periodic solutions of the ordinary differential
equations describing the nonlinear vertical oscillations.  The left
panel of Fig.~\ref{f:h} shows azimuthal profiles of $H$ for
eccentricities up to $0.6$, by which point an extreme compression has
occurred near pericentre.  The right panel illustrates an almost
universal behaviour near pericentre for larger eccentricities, when
the variables are rescaled in terms of the minimum value of $H$, which
is absurdly small in the case $e=0.75$.  In this regime vertical
gravity is unimportant near pericentre; its variation around the orbit
does however induce a dynamical collapse that bounces near pericentre
because of the large pressure that develops there.

\begin{figure}
\subfigure{\includegraphics[trim=0cm 0cm 0cm 0cm, clip=true,width=0.44\textwidth]{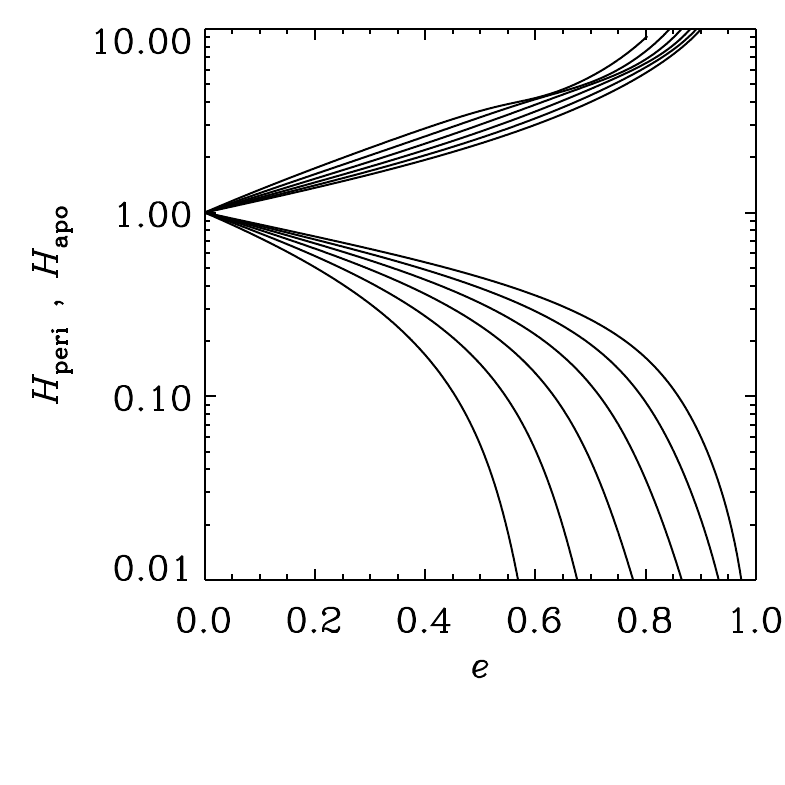} }
\caption{Minimum and maximum scaleheight, in units of the hydrostatic
  value for a circular disc with the same semi-latus rectum, for
  eccentric discs with no eccentricity gradient and with $\gamma=1$,
  $1.2$, $1.4$, $1.6$, $1.8$ and $2$.  The outermost curves are for
  $\gamma=1$ and the innermost ones for $\gamma=2$.}
\label{f:hperi_hapo}
\end{figure}

Fig.~\ref{f:hperi_hapo} shows how the minimum and maximum values of
$H$, which are obtained at pericentre and apocentre respectively,
depend on $e$ for various $\gamma$ in the absence of an eccentricity
gradient.  The extreme behaviour occurs at larger $e$ when $\gamma$ is
larger; this can be understood as an extension of the linear result
that $\tilde H=-3E/\gamma$ when $E'=0$.

\begin{figure}
\subfigure{\includegraphics[trim=0cm 0cm 0cm 0cm, clip=true,width=0.44\textwidth]{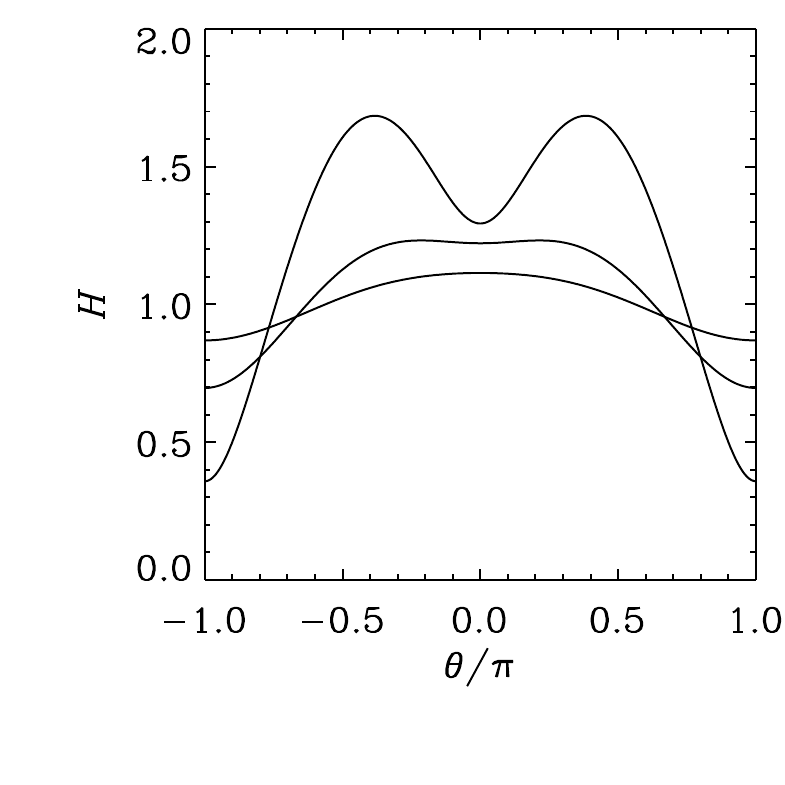} }
\caption{Laminar flows for eccentric discs with $\gamma=5/3$ and with
  eccentricity gradients $\lambda e'=0.3$, $0.6$ and $0.9$, but with a
  circular reference orbit ($e=0$) and no twist ($\omega'=0$).  For
  $\lambda e'=0.3$ the scaleheight varies sinusoidally with a maximum
  at pericentre.  For larger eccentricity gradients the profile
  develops a local minimum at pericentre and two maxima elsewhere, but
  the behaviour is not extreme.}
\label{f:h_lep}
\end{figure}

Finally, Fig.~\ref{f:h_lep} shows some laminar flows in discs with an
eccentricity gradient but with a circular reference orbit.  Here the
vertical oscillation is driven only by the orbital velocity divergence
and occurs only if $\gamma\ne1$.  The behaviour is qualitatively
different from that in Fig.~\ref{f:h} and the oscillations are of more
modest amplitude.

\subsection{Comparison with \citet{2001MNRAS.325..231O}}

If heating and cooling are neglected and the stress is an
instantaneous bulk viscous stress, then the relevant parts of
equations (211)--(217) of \citet{2001MNRAS.325..231O} reduce to
\begin{equation}
  (1+e\cos\theta)^2\p_\theta\ln f_2=-(\gamma+1)f_3-(\gamma-1)g_1,
\end{equation}
\begin{equation}
  (1+e\cos\theta)^2\p_\theta f_3=-f_3^2-(1+e\cos\theta)^3+f_2[1-\alpha_\rmb(g_1+f_3)].
\end{equation}
The correspondence with the equations of this section is
$f_2\mapsto\hat p/\hat\rho H^2$, $f_3\mapsto w(GM/\lambda^3)^{-1/2}$,
$g_1\mapsto\Delta(GM/\lambda^3)^{-1/2}$.

\section{Conclusion}
\label{s:conclusion}

Although Keplerian discs are usually assumed to be circular, the
general Keplerian disc is elliptical, in accordance with Kepler's
first law.  As found by \citet{1992MNRAS.255...92S} and other authors,
viscosity (and other dissipative effects) do not necessarily lead to
the circularization of a disc, even though circular orbits have the
least energy for a given angular momentum; this is because dissipative
forces can tap the reservoir of orbital energy by causing mass
redistribution.  Eccentricity may result from the initial conditions
of the disc (as in the case of a disc formed through tidal disruption
of a body on an elliptical orbit), from secular forcing by a companion
with an elliptical orbit, from resonant forcing by a companion with a
circular or elliptical orbit, or from instability of a circular disc.

In this paper we have revisited the theory of eccentric discs
developed by \citet{2001MNRAS.325..231O}.  In particular, we have
formulated a local model, which is a generalization of the well known
shearing sheet (or box) to the geometry of an eccentric disc.  We have
discussed the simplest hydrodynamic solutions in the local model,
which are necessarily non-hydrostatic and involve a vertical
oscillation at the orbital period.  These oscillations can become
highly nonlinear and exhibit extreme behaviour at eccentricities
significantly less than unity, especially if the disc behaves
isothermally.  It would be valuable to determine, using numerical
simulations, whether these extreme solutions are realized in practice.
We have also computed the stresses associated with the laminar flows
in a linear regime and derived the associated global evolutionary
equation for the eccentricity, which differs significantly from a
two-dimensional theory that neglects the vertical structure and
oscillation of the disc.

A question not addressed in this paper is a possible vertical
dependence of the eccentricity.  In the absence of viscosity,
turbulence and magnetic fields, layers of the disc at different
heights are relatively weakly coupled by pressure gradients and can
undergo independent epicyclic oscillations to some extent.  As
discussed by \citet{2006MNRAS.372.1829L}, this allows the eccentricity
to propagate radially with a non-trivial vertical profile.  There may
therefore be a transition in behaviour when the viscous, turbulent or
magnetic stresses are very small.

In the companion paper \citep{2014BO} we use the local model to
analyse the linear hydrodynamic stability of an eccentric disc.  In
the absence of viscosity and magnetic fields, eccentric discs are
susceptible to a hydrodynamic instability that excites internal
(inertial) waves and may induce hydrodynamic turbulence.  This is
likely to be important for the evolution of the eccentricity, but also
potentially for transport processes and mixing.

\section*{Acknowledgements}

This research was supported by STFC through grants ST/J001570/1 and
ST/L000636/1.

\appendix

\section{Geometrical relations}
\label{s:relations}

From the polar equation~(\ref{ellipse}) of an ellipse,  we have
\begin{equation}
  \f{\lambda}{R}=1+e\cos(\phi-\omega)
\label{lor}
\end{equation}
and therefore
\begin{equation}
  (\p_\phi^2+1)\left(\f{\lambda}{R}\right)=1,
\end{equation}
which implies
\begin{equation}
  R^2+2R_\phi^2-RR_{\phi\phi}=\f{R^3}{\lambda}
\label{rpp}
\end{equation}
and simplifies the expression for
\begin{equation}
  \Gamma^\lambda_{\phi\phi}=-\f{R^2}{\lambda R_\lambda}.
\end{equation}
By differentiating equation~(\ref{lor}) once with respect to $\phi$,
we find
\begin{equation}
  \f{\lambda}{R^2}(R-\rmi R_\phi)=1+E\,\rme^{-\rmi\phi},
\label{rp}
\end{equation}
which can also be written as
\begin{equation}
  \p_\phi(R\,\rme^{\rmi\phi})=\f{\rmi R^2}{\lambda}(\rme^{\rmi\phi}+E)
\end{equation}
or
\begin{equation}
  -\p_\phi(R\,\rme^{\rmi\phi})^{-1}=\f{\rmi}{\lambda}(\rme^{-\rmi\phi}+E\,\rme^{-2\rmi\phi}).
\end{equation} 
Differentiating with respect to $\lambda$ and interchanging the order
of differentiation, we find
\begin{equation}
  \p_\phi\left(\f{R_\lambda}{R^2\,\rme^{\rmi\phi}}\right)=-\f{\rmi}{\lambda^2}\left[\rme^{-\rmi\phi}+(E-\lambda E')\,\rme^{-2\rmi\phi}\right].
\end{equation}
This in turn implies
\begin{equation}
  \p_\phi\left(\f{R^2\,\rme^{\rmi\phi}}{R_\lambda}\right)=\f{\rmi R^4}{\lambda^2R_\lambda^2}(\rme^{\rmi\phi}+E-\lambda E').
\label{rlp}
\end{equation}

\section{Expressions for the coefficients}
\label{s:coefficients}

Let $\theta=\varphi(t)-\omega$ be the true anomaly on the reference
orbit.  Then the following quantities, when evaluated at the reference
point as described above, may be written explicitly in terms of
$\theta$:
\begin{equation}
  R=\f{\lambda}{1+e\cos\theta},
\end{equation}
\begin{equation}
  R_\lambda=\f{1+(e-\lambda e')\cos\theta-\lambda e\omega'\sin\theta}{(1+e\cos\theta)^2},
\end{equation}
\begin{equation}
  R_\phi=\f{\lambda e\sin\theta}{(1+e\cos\theta)^2},
\end{equation}
\begin{equation}
  J=\f{\lambda[1+(e-\lambda e')\cos\theta-\lambda e\omega'\sin\theta]}{(1+e\cos\theta)^3},
\end{equation}
\begin{equation}
  \Omega=\left(\f{GM}{\lambda^3}\right)^{1/2}(1+e\cos\theta)^2,
\end{equation}
\begin{eqnarray}
  \lefteqn{\lambda\Omega_\lambda=\left(\f{GM}{\lambda^3}\right)^{1/2}\bigg[-\f{3}{2}(1+e\cos\theta)^2}&\nonumber\\
  &&+2(1+e\cos\theta)(\lambda e'\cos\theta+\lambda e\omega'\sin\theta)\bigg],
\end{eqnarray}
\begin{equation}
  \Omega_\phi=\left(\f{GM}{\lambda^3}\right)^{1/2}(-2e\sin\theta)(1+e\cos\theta),
\end{equation}
\begin{equation}
  \Phi_2=\left(\f{GM}{\lambda^3}\right)(1+e\cos\theta)^3,
\end{equation}
\begin{equation}
  \Delta=\left(\f{GM}{\lambda^3}\right)^{1/2}\f{(1+e\cos\theta)[\lambda e'\sin\theta-\lambda e\omega'(\cos\theta+e)]}{1+(e-\lambda e')\cos\theta-\lambda e\omega'\sin\theta},
\end{equation}
\begin{equation}
  g_{\lambda\lambda}=\f{[1+(e-\lambda e')\cos\theta-\lambda e\omega'\sin\theta]^2}{(1+e\cos\theta)^4}
\end{equation}
\begin{equation}
  \lambda^{-1}g_{\lambda\phi}=\f{[1+(e-\lambda e')\cos\theta-\lambda e\omega'\sin\theta]e\sin\theta}{(1+e\cos\theta)^4},
\end{equation}
\begin{equation}
  \lambda^{-2}g_{\phi\phi}=\f{1+2e\cos\theta+e^2}{(1+e\cos\theta)^4},
\end{equation}
\begin{equation}
  g^{\lambda\lambda}=\f{(1+e\cos\theta)^2(1+2e\cos\theta+e^2)}{[1+(e-\lambda e')\cos\theta-\lambda e\omega'\sin\theta]^2},
\end{equation}
\begin{equation}
  \lambda g^{\lambda\phi}=-\f{(1+e\cos\theta)^2e\sin\theta}{1+(e-\lambda e')\cos\theta-\lambda e\omega'\sin\theta},
\end{equation}
\begin{equation}
  \lambda^2g^{\phi\phi}=(1+e\cos\theta)^2,
\end{equation}
\begin{equation}
  \Gamma^\lambda_{\lambda\phi}=\f{\lambda e'\sin\theta-\lambda e\omega'(\cos\theta+e)}{(1+e\cos\theta)[1+(e-\lambda e')\cos\theta-\lambda e\omega'\sin\theta]},
\end{equation}
\begin{equation}
  \lambda^{-1}\Gamma^\lambda_{\phi\phi}=-\f{1}{1+(e-\lambda e')\cos\theta-\lambda e\omega'\sin\theta},
\end{equation}
\begin{equation}
  \lambda\Gamma^\phi_{\lambda\phi}=\f{1+(e-\lambda e')\cos\theta-\lambda e\omega'\sin\theta}{1+e\cos\theta},
\end{equation}
\begin{equation}
  \Gamma^\phi_{\phi\phi}=\f{2e\sin\theta}{1+e\cos\theta}.
\end{equation}
On the right-hand sides of these equations, quantities such as
$\lambda$, $e$, $e'$ and $\omega'$ are to be evaluated on the
reference orbit $\lambda=\lambda_0$.  Note that the combinations of
metric and connection components with various powers of $\lambda$ make
these quantities dimensionless.  Note also that the orbital velocity
divergence $\Delta$ is non-zero when the eccentricity gradient $E'$ is
non-zero.

It is possible, in principle, to write these coefficients in terms of
$\tau$ rather than $\theta$.  These quantities can be related through
Kepler's equation and the mean and eccentric anomalies, or by
inverting the function $\varphi(\tau)$.  However, in practice, it is
easier to use $\theta$ instead of $\tau$ as a time-like variable.
Time-derivatives of these quantities at the reference point may be
evaluated using the rule $\p_\tau=\Omega\p_\theta$.

\section{Magnetohydrodynamic equations}
\label{s:mhd}

In ideal magnetohydrodynamics (MHD) the equation of motion can be written
in the form
\begin{eqnarray}
  \lefteqn{(\p_t+u^j\nabla_j)u^i=-g^{ij}\left[\nabla_j\Phi+\f{1}{\rho}\nabla_j\left(p+\f{B^2}{2\mu_0}\right)\right]}&\nonumber\\
  &&+\f{1}{\mu_0}B^j\nabla_jB^i,
\end{eqnarray}
and the induction equation in the form
\begin{equation}
  (\p_t+u^j\nabla_j)B^i=B^j\nabla_ju^i-B^i\nabla_ju^j,
\end{equation}
while the solenoidal condition is
\begin{equation}
  \nabla_iB^i=0.
\end{equation}
In the local model using non-shearing coordinates, under similar
scaling assumptions to those used in deriving the hydrodynamic
equations, the three components of the equation of motion
(\ref{dvxi})--(\ref{dvzeta}) are therefore modified to
\begin{eqnarray}
  \lefteqn{\rmD v^\xi+2\Gamma^\lambda_{\lambda\phi}\Omega v^\xi+2\Gamma^\lambda_{\phi\phi}\Omega v^\eta}&\nonumber\\
  &&=-\f{1}{\rho}\left[g^{\lambda\lambda}\p_\xi\left(p+\f{B^2}{2\mu_0}\right)+g^{\lambda\phi}\p_\eta\left(p+\f{B^2}{2\mu_0}\right)\right]\nonumber\\
  &&+\f{1}{\rho\mu_0}(B^\xi\p_\xi+B^\eta\p_\eta+B^\zeta\p_\zeta)B^\xi,
\end{eqnarray}
\begin{eqnarray}
  \lefteqn{\rmD v^\eta+(\Omega_\lambda+2\Gamma^\phi_{\lambda\phi}\Omega)v^\xi+(\Omega_\phi+2\Gamma^\phi_{\phi\phi}\Omega)v^\eta}&\nonumber\\
  &&=-\f{1}{\rho}\left[g^{\lambda\phi}\p_\xi\left(p+\f{B^2}{2\mu_0}\right)+g^{\phi\phi}\p_\eta\left(p+\f{B^2}{2\mu_0}\right)\right]\nonumber\\
  &&+\f{1}{\rho\mu_0}(B^\xi\p_\xi+B^\eta\p_\eta+B^\zeta\p_\zeta)B^\eta,
\end{eqnarray}
\begin{eqnarray}
  \lefteqn{\rmD v^\zeta=-\Phi_2\zeta-\f{1}{\rho}\p_\zeta\left(p+\f{B^2}{2\mu_0}\right)}&\nonumber\\
  &&+\f{1}{\rho\mu_0}(B^\xi\p_\xi+B^\eta\p_\eta+B^\zeta\p_\zeta)B^\zeta,
\end{eqnarray}
while the three components of the induction equation are
\begin{eqnarray}
  \lefteqn{\rmD B^\xi=(B^\xi\p_\xi+B^\eta\p_\eta+B^\zeta\p_\zeta)v^\xi}&\nonumber\\
  &&-B^\xi(\Delta+\p_\xi v^\xi+\p_\eta v^\eta+\p_\zeta v^\zeta),
\end{eqnarray}
\begin{eqnarray}
  \lefteqn{\rmD B^\eta=\Omega_\lambda B^\xi+\Omega_\phi B^\eta+(B^\xi\p_\xi+B^\eta\p_\eta+B^\zeta\p_\zeta)v^\eta}&\nonumber\\
  &&-B^\eta(\Delta+\p_\xi v^\xi+\p_\eta v^\eta+\p_\zeta v^\zeta),
\end{eqnarray}
\begin{eqnarray}
  \lefteqn{\rmD B^\zeta=(B^\xi\p_\xi+B^\eta\p_\eta+B^\zeta\p_\zeta)v^\zeta}&\nonumber\\
  &&-B^\zeta(\Delta+\p_\xi v^\xi+\p_\eta v^\eta+\p_\zeta v^\zeta),
\end{eqnarray}
and the solenoidal condition is
\begin{equation}
  \p_\xi B^\xi+\p_\eta B^\eta+\p_\zeta B^\zeta=0.
\end{equation}

Under the transformation to shearing coordinates these equations become
\begin{eqnarray}
  \lefteqn{\rmD v^\xi+2\Gamma^\lambda_{\lambda\phi}\Omega v^\xi+2\Gamma^\lambda_{\phi\phi}\Omega v^\eta}&\nonumber\\
  &&=-\f{1}{\rho}\left[g^{\lambda\lambda}(\p_\xi'+\beta\p_\eta')\left(p+\f{B^2}{2\mu_0}\right)+g^{\lambda\phi}\alpha\p_\eta'\left(p+\f{B^2}{2\mu_0}\right)\right]\nonumber\\
  &&+\f{1}{\rho\mu_0}[B^\xi(\p_\xi'+\beta\p_\eta')+B^\eta\alpha\p_\eta'+B^\zeta\p_\zeta']B^\xi,
\end{eqnarray}
\begin{eqnarray}
  \lefteqn{\rmD v^\eta+(\Omega_\lambda+2\Gamma^\phi_{\lambda\phi}\Omega)v^\xi+(\Omega_\phi+2\Gamma^\phi_{\phi\phi}\Omega)v^\eta}&\nonumber\\
  &&=-\f{1}{\rho}\left[g^{\lambda\phi}(\p_\xi'+\beta\p_\eta')\left(p+\f{B^2}{2\mu_0}\right)+g^{\phi\phi}\alpha\p_\eta'\left(p+\f{B^2}{2\mu_0}\right)\right]\nonumber\\
  &&+\f{1}{\rho\mu_0}[B^\xi(\p_\xi'+\beta\p_\eta')+B^\eta\alpha\p_\eta'+B^\zeta\p_\zeta']B^\eta,
\end{eqnarray}
\begin{eqnarray}
  \lefteqn{\rmD v^\zeta=-\Phi_2\zeta-\f{1}{\rho}\p_\zeta'\left(p+\f{B^2}{2\mu_0}\right)}&\nonumber\\
  &&+\f{1}{\rho\mu_0}[B^\xi(\p_\xi'+\beta\p_\eta')+B^\eta\alpha\p_\eta'+B^\zeta\p_\zeta']B^\zeta,
\end{eqnarray}
\begin{eqnarray}
  \lefteqn{\rmD B^\xi=[B^\xi(\p_\xi'+\beta\p_\eta')+B^\eta\alpha\p_\eta'+B^\zeta\p_\zeta']v^\xi}&\nonumber\\
  &&-B^\xi[\Delta+(\p_\xi'+\beta\p_\eta')v^\xi+\alpha\p_\eta'v^\eta+\p_\zeta'v^\zeta],
\end{eqnarray}
\begin{eqnarray}
  \lefteqn{\rmD B^\eta=\Omega_\lambda B^\xi+\Omega_\phi B^\eta+[B^\xi(\p_\xi'+\beta\p_\eta')+B^\eta\alpha\p_\eta'+B^\zeta\p_\zeta']v^\eta}&\nonumber\\
  &&-B^\eta[\Delta+(\p_\xi'+\beta\p_\eta')v^\xi+\alpha\p_\eta'v^\eta+\p_\zeta'v^\zeta],
\end{eqnarray}
\begin{eqnarray}                                                                  \lefteqn{\rmD B^\zeta=[B^\xi(\p_\xi'+\beta\p_\eta')+B^\eta\alpha\p_\eta'+B^\zeta\p_\zeta']v^\zeta}&\nonumber\\
  &&-B^\zeta[\Delta+(\p_\xi'+\beta\p_\eta')v^\xi+\alpha\p_\eta'v^\eta+\p_\zeta'v^\zeta],
\end{eqnarray}
\begin{equation}
  (\p_\xi'+\beta\p_\eta')B^\xi+\alpha\p_\eta'B^\eta+\p_\zeta'B^\zeta=0.
\end{equation}

\label{lastpage}

\end{document}